\documentclass[fleqn,usenatbib]{mnras}

\usepackage{newtxtext,newtxmath}
\usepackage[T1]{fontenc}

\DeclareRobustCommand{\VAN}[3]{#2}
\let\VANthebibliography\thebibliography
\def\thebibliography{\DeclareRobustCommand{\VAN}[3]{##3}\VANthebibliography}

\usepackage{graphicx}	% Including figure files
\usepackage{amsmath}	% Advanced maths commands
\newcommand{\avg}[1]{\left\langle#1\right\rangle}

\newif\iftrack
\iftrack
\newcommand{\added}[1]{{\bf #1}}
\newcommand{\deleted}[1]{}
\newcommand{\replaced}[2]{{\bf #2}}
\else
\newcommand{\added}[1]{{#1}}
\newcommand{\deleted}[1]{}
\newcommand{\replaced}[2]{{#2}}
\fi
\newif\iftracktwo
\iftracktwo
\newcommand{\addedtwo}[1]{{\bf #1}}
\newcommand{\deletedtwo}[1]{}
\newcommand{\replacedtwo}[2]{{\bf #2}}
\else
\newcommand{\addedtwo}[1]{{#1}}
\newcommand{\deletedtwo}[1]{}
\newcommand{\replacedtwo}[2]{{#2}}
\fi

\title[Wavelet scattering transform and LIM]{Exploration of 3D wavelet scattering transform coefficients for line-intensity mapping measurements}

\author[D. T. Chung]{
Dongwoo T.~Chung$^{1,2}$\thanks{E-mail: dongwooc@cita.utoronto.ca}
\\
$^{1}$Canadian Institute for Theoretical Astrophysics, University of Toronto, 60 St. George Street, Toronto, ON M5S 3H8, Canada\\
$^{2}$Dunlap Institute for Astronomy and Astrophysics, University of Toronto, 50 St. George Street, Toronto, ON M5S 3H4, Canada
}

\date{Accepted XXX. Received YYY; in original form ZZZ}

\pubyear{2022}

\begin{document}
\label{firstpage}
\pagerange{\pageref{firstpage}--\pageref{lastpage}}
\maketitle

\begin{abstract}
The wavelet scattering transform (WST) has recently gained attention in the context of large-scale structure studies, being a possible generator of summary statistics encapsulating non-Gaussianities beyond the reach of the conventional power spectrum. This work examines the three-dimensional solid harmonic WST in the context of a three-dimensional line-intensity mapping measurement to be undertaken by current and proposed phases of the CO Mapping Array Project (COMAP). The WST coefficients demonstrate interpretable behaviour in the context of noiseless CO line-intensity simulations. The contribution of the cosmological $z\sim3$ signal to these coefficients is also detectable in principle even in the Pathfinder phase of COMAP. \replacedtwo{In Fisher forecasts based on}{Using the peak-patch method to generate} large numbers of simulations \replacedtwo{that incorporate}{and incorporating} observational noise, \addedtwo{we numerically estimate covariance matrices and show that careful choices of WST hyperparameters and rescaled or reduced coefficient sets are both necessary to keep covariances well-conditioned. Fisher forecasts show that} even a reduced `shapeless' set of $\ell$-averaged WST coefficients show constraining power that can exceed that of the power spectrum alone even with similar detection significance. The full WST could improve parameter constraints even over the combination of the power spectrum and the voxel intensity distribution, showing that it uniquely encapsulates shape information about the line-intensity field. However, practical applications urgently require further understanding of the WST in key contexts like covariances and cross-correlations.
\end{abstract}


%%%%%%%%%%%%%%%%%%%%%%%%%%%%%%%%%%%%%%%%%%%%%%%%%%

%%%%%%%%%%%%%%%%% BODY OF PAPER %%%%%%%%%%%%%%%%%%

\section{Introduction}
\label{sec:intro}
The field of line-intensity mapping (LIM) promises to chart the large-scale structure (LSS) of the Universe as traced by a variety of atomic and molecular spectral lines, and thereby statistically and efficiently constrain high-redshift astrophysics and cosmology (cf.~\citealt{LIM2017,LIM2019} and~\citealt{BernalKovetz22} for comprehensive reviews of the field in recent years). LIM experiments will observe large cosmological volumes for thousands of hours and obtain spatial-spectral data cubes that contain -- underneath intimidating layers of noise and other contaminants -- line emission from every galaxy in the survey volume, faint or bright. The success of LIM science then depends on leveraging fluctuations in this integrated line emission with appropriate statistical methods.

The conventional approach in both simulations and analyses has been to use the spherically averaged three-dimensional power spectrum (or $P(k)$, a function of comoving wavenumber $k$) as the principal LIM summary statistic. But the processes of structure formation and galaxy formation, and thus the line-intensity map produced by these processes, have significant non-Gaussianities to which the power spectrum is insensitive. Previous works, starting no later than~\cite{Breysse17}, have proposed and developed the voxel intensity distribution (VID) as a supplementary LIM summary statistic that is sensitive to these non-Gaussianities.

Outside of LIM contexts, the need for statistical methods that can help extract non-Gaussian information has led to the use of approaches originally developed in computer vision contexts. Works of ambitious scale (e.g.,~\citealt{CAMELSMultifield}) have explored the use of convolutional neural networks, which in essence apply repeated filtering, downsampling, and clipping operations to the input. The resulting multi-scale, nonlinear nature of the operation strongly motivates use of these networks in LSS studies. However, use of such a network first and foremost requires determination of its weights, typically by optimising performance on a training set as measured by a loss function.

The need for a sizeable and representative training set is problematic in the context of high-redshift astrophysics and cosmology. Prior information at high redshift may well be both incredibly ill-determined and constantly evolving, demanding repeated, extensive, possibly expensive re-training of the network weights. \cite{Pfeffer19} considered convolutional neural networks in the context of LIM with carbon monoxide (CO) line emission, and demonstrated strong degradation of the accuracy of inferred astrophysical constraints from unanticipated signal or noise effects. Such degradation reflects a tenuous interpretability (and credibility) of any outputs from these kinds of networks, strongly dependent on the training set and its limitations.

In view of such concerns, the wavelet scattering transform (WST) has recently gained attention in the LSS literature. In comparison to convolutional neural networks, the WST (based on principles developed by~\citealt{Mallat12}) has a similar multi-scale, multi-layer nature suiting it to cosmology applications, but potentially does not sacrifice interpretability to the same extent.\added{ \cite{Cheng20} applied the WST to a simulated cosmological observable for the first time, specifically in the context of weak lensing convergence maps. That work suggested strong constraining power and interpretability compared to both traditional estimators -- namely the power spectrum and peak counts -- and convolutional neural networks. Further exploration by~\cite{Cheng21a} explicitly compared the WST to using the bispectrum, a traditional higher-order statistic that does capture non-Gaussian information, and found the WST to still be more robust and constraining while also being more Gaussian in variance.

} This promise has already motivated exploration of the two-dimensional WST in the context of 21 cm cosmology~\citep{Greig22}, as well as of the three-dimensional WST in the context of simulated matter density fields~\citep{Valogiannis22a} and real-world galaxy surveys~\citep{Valogiannis22b}. However, work has yet to examine the application of three-dimensional WST coefficients (in the manner of~\citealt{Valogiannis22b,Valogiannis22a}) to noisy observations (as~\citealt{Greig22} did but for a two-dimensional WST implementation).

This paper aims to examine the three-dimensional spherical harmonic WST in the context of the CO Mapping Array Project (COMAP;~\citealt{Cleary21}), and in particular its measurement of the clustering of CO(1--0) line emission at $z\sim3$ (although COMAP ultimately aims to also map CO emission at $z\sim7$). In particular, we will explore the following questions:
\begin{itemize}
    \item What is the detectability of cosmological CO emission in the WST coefficients derived from a simulated COMAP observation?
    \item Do parameter constraints from these coefficients improve significantly over the previously studied combination of $P(k)$ with the VID?
\end{itemize}

The paper's organisation is as follows. We will review the summary statistics to be calculated, including the WST, in~\autoref{sec:stats}, then consider in~\autoref{sec:methods} the methods for forecasting inferences using those summary statistics derived from simulated observations. We will examine the resulting forecasts in~\autoref{sec:results}, before discussing implications of these results in~\autoref{sec:discussion} and concluding in~\autoref{sec:conclusions}.

Unless otherwise stated, we assume base-10 logarithms, and a $\Lambda$CDM cosmology with parameters $\Omega_m = 0.286$, $\Omega_\Lambda = 0.714$, $\Omega_b =0.047$, $H_0=100h$\,km\,s$^{-1}$\,Mpc$^{-1}$ with $h=0.7$, $\sigma_8 =0.82$, and $n_s =0.96$, to maintain consistency with previous simulations used by~\cite{Ihle19}. Distances carry an implicit $h^{-1}$ dependence throughout, which propagates through masses (all based on virial halo masses, proportional to $h^{-1}$) and volume densities ($\propto h^3$).

\section{Statistics}
\label{sec:stats}

This section will provide an overview of the three summary statistics applied to our simulated observations: the power spectrum (\autoref{sec:pspec}), the VID (\autoref{sec:vid}), and the WST (\autoref{sec:wst}).
\subsection{Power spectrum}
\label{sec:pspec}
The full three-dimensional power spectrum of a field $f(\mathbfit{x})$, expressed as a function of a three-dimensional positional vector $\mathbfit{x}$, is given through its Fourier transform $\tilde{f}(\mathbfit{k})$, a function of the wavevector $\mathbfit{k}$:
\begin{equation}
    P(\mathbfit{k}) \propto |\tilde{f}(\mathbfit{k})|^2 = \tilde{f}^*(\mathbfit{k})\tilde{f}(\mathbfit{k}),
\end{equation}
with the proportionality constant being the inverse of the survey volume. The spherically-averaged three-dimensional power spectrum $P(k)$ is then a binned reduction of this $P(\mathbfit{k})$ in spherical shells of constant wavenumber $k$:
\begin{equation}
    P(k) = \avg{P(\mathbfit{k})}_{|\mathbfit{k}|=k}.
\end{equation}
Because the Universe is \emph{mostly} isotropic and homogeneous on large cosmological scales, $P(k)$, the monopole of the power spectrum, contains most of the information that would be contained in $\tilde{f}(\mathbfit{k})$ or $P(\mathbfit{k})$, and has the advantage that we can average independently observed Fourier modes across $k$-shells to obtain a better measurement of $P(k)$. The result is a measurement of the power of field contrast fluctuations associated with different comoving scales $L\sim 2\pi/k$.

Ultimately, the fact that we observe LSS in redshift space and not in real comoving space does introduce anisotropies, which are captured in higher-order multipoles of $P(\mathbfit{k})$. But the plain monopole $P(k)$ remains a veritable workhorse of basic cosmology theory. We commonly describe the distribution of matter across the Universe with the matter power spectrum $P_m(k)$. We then describe the density contrast of a biased tracer of matter with a power spectrum $b^2P_m(k)$, given some linear tracer bias $b$ by which the matter density contrast $\delta_m$ scales to the tracer density contrast $\delta = b\delta_m$. If using a dark matter halo population as a tracer, the halo bias $b(M_h)$ as a function of halo (virial) mass $M_h$ is well studied (cf.~\citealt{Manera10,Pillepich10,Tinker10}). The bias of galaxies or other associated tracers, including the line intensity contrast in a LIM observation, then follows from the galaxy--halo connection~\citep{WechslerTinker18}. Redshift-space distortions, theories of nonlinear bias, and other complications (as considered by, e.g.,~\citealt{Bernal19,BBK19} or~\citealt{MD22}) operate on top of these fundamental ideas.

Despite the loss of non-Gaussian information and anisotropic information, the greatest strength of $P(k)$ as a summary statistic is in the ease of understanding its variance and covariance. Power spectra of statistically independent signal and noise components simply add; the variance of the total $P(k)$ for a given $k$-shell containing $N_m(k)$ Fourier modes is simply $P^2(k)/{N_m(k)}$; and the covariance matrix of $P(k)$ (especially in a noise-dominated observation) is largely diagonal\footnote{The covariance in practice will not be diagonal as observational effects could certainly introduce mode coupling. Furthermore, off-diagonal terms absolutely must be considered against higher-order multipoles of $P(\mathbfit{k})$.}, easing statistical considerations in measurement and inferences.

Finally, it is straightforward to consider the cross-correlation power spectrum between two fields $f(\mathbfit{x})$ and $g(\mathbfit{x})$ through their Fourier transforms $\tilde{f}(\mathbfit{k})$ and $\tilde{g}(\mathbfit{k})$:
\begin{equation}
    P_{f\times g}(\mathbfit{k})\propto \tilde{f}^*(\mathbfit{k})\tilde{g}(\mathbfit{k}),
\end{equation}
with the proportionality constant and the spherically-averaged $P_{f\times g}(k)$ obtained in the same way as for the auto-correlation power spectrum. In fact, it is fair to say that the auto-correlation $P(k)$ is merely a special case of the cross power spectrum where $f(x) = g(x)$. Given the complexity of galaxy formation and the need for multi-tracer astrophysics on cosmological scales through different LIM and LSS observations, not to mention the need for mitigation of different foregrounds and systematics present in these observations, cross-correlations between different experiments surveying common comoving volumes are highly desirable. At present, the cross power spectrum remains the best studied and thus best understood statistic to address these needs.
\subsection{VID}
\label{sec:vid}
The VID was first put forward in LIM contexts by~\cite{Breysse17}, who characterise it as an extension of the $\mathcal{P}(D)$ technique in radio astronomy put forward by~\cite{Scheuer57}. The central idea is to characterise the distribution of intensities (or, in the time and language of~\cite{Scheuer57}, deflexions on a power recording chart---hence $\mathcal{P}(D)$ for probability of deflexion) in a confused map of observed emission from a population of sources, and relate this to the underlying luminosity function of sources. LIM experiments tend to operate in a confused regime to focus on resolving the large-scale clustering of line emission, so the VID technique holds extreme relevance. The study of~\cite{Breysse17}, as well as follow-up by~\cite{Ihle19} among others, showed that the VID truly encapsulates non-Gaussian, small-scale information beyond the power spectrum, related to the shape of the source luminosity function and to the stochasticity of line emission.

Being a histogram of voxel intensities (or temperatures, in the convention of a cm-wavelength experiment like COMAP) with some bin width $\Delta T$ and range $(T_\text{min},T_\text{max})$, the VID is straightforward to calculate from any data cube. Its interpretation, however, is significantly less straightforward due the nature of the VID of a confused map. The VID is given by a series of recursive convoluted probability distributions of voxel intensities conditioned on voxels containing specific numbers of emitters, complicated by the clustering of sources. Observational effects like instrumental resolution and line widths further affect the VID. Nonetheless, the behaviour of the VID -- including its covariance -- can still be understood through simulations as in~\cite{Ihle19}, and even through concrete mathematical expressions~\citep{SatoPolitoBernal22}.

Furthermore, while not as straightforwardly as for the power spectrum, one \emph{can} design one-point cross-correlation statistics between multiple tracers to reject disjoint contaminants. \cite{Breysse19} propose a one-point analysis for a 21 cm intensity map with VIDs conditioned on the presence or absence of a galaxy in an external catalogue surveying the same volume, showing that the ratio of Fourier transforms of these conditional VIDs provides a one-point cross-correlation statistic unbiased by the presence of foregrounds. Similar joint VID analyses can be applied in the context of LIM observations and internal or external cross-correlations between LIM datasets (Breysse et al., in prep.).

\subsection{Solid harmonic WST}
\label{sec:wst}
As with $P(k)$ and the VID, the WST has a concrete mathematical description and strong interpretability, reviewed pedagogically (albeit in a chiefly two-dimensional context) by~\cite{Cheng21}. The essence is to convolve the input field with a family of wavelets, and the complex modulus of this convolved field is the `scattered' output. The coefficients of the WST come from spatial averages of the output after different levels of scattering. The resemblance of the WST to convolutional neural networks arises from the possibility of recursive application of this scattering operation (including the nonlinear `clipping' introduced by the modulus operation and the downsampling coming from averaging the scattered outputs) with wavelets of different scales and anisotropies.

In practice, works exploring the WST only deal with the first- and second-order scattering coefficients, i.e., the field average after one or two scattering operations. This is already sufficient to probe scale interactions, in ways that depend on the wavelets used and their frequency response.

Following the implementation of the three-dimensional WST by~\cite{kymatio} in the Kymatio software package, we specifically will consider a family of solid harmonic wavelets. This solid harmonic WST was originally considered by~\cite{Eickenberg18}\footnote{See also~\cite{Eickenberg17} for a similar overview of the solid harmonic WST in its original motivating context.} in the context of molecular energy regression. Molecular properties like the ground state energy depend on the electronic density, popularly calculated via density-functional theory. Although this is not necessarily a prohibitive calculation, the atomistic state -- i.e., the description of a molecule in terms of the constituent atoms and their positions -- is still a simpler description than the true electronic density, and can be geometrically represented through different pseudo-electronic densities. The aim of~\cite{Eickenberg18} was to obtain multilinear regressions to map the WST of this molecular geometry directly to various molecular properties in a predictive fashion, which the work demonstrated with excellent performance metrics.

In both the molecular energy regression problem and our case study, the goal is to constrain parameters describing the structure underlying the input field, be it an organic molecule or a sample of high-redshift galaxies. However, in our case the starting point is not an approximate, noise-free description of particle density, but rather an empirical, noisy measurement of true CO volume emissivity. Nonetheless, the central idea behind the WST as a summary statistic probing multi-scale interactions remains in all contexts, and even at the time of the work of~\cite{Eickenberg18}, solid harmonic wavelets had already been used in three-dimensional image processing contexts outside of quantum chemistry\footnote{Specifically,~\cite{Eickenberg18} provide the example of~\cite{Reisert09}, whose work considers harmonic filters in the context of three-dimensional feature detection, with a demonstrative example using microscopic imagery of pollen grains.}.

The solid harmonic wavelets are indexed by the logarithmic scale parameter $j$, the angular frequency index $\ell$, and the orientation index $m\in[-\ell,\ell]$. Furthermore, we will express these wavelets as a function of the position vector $\mathbfit{u}$ in image coordinates (so in units of pixels or grid width) rather than physical or comoving coordinates, to mirror the description given in other works.

Before introducing the scale $j$, the `mother' wavelets are given by the Laplacian spherical harmonic $Y_\ell^m$ rescaled by $|u|^\ell$ to make a (regular) solid harmonic, then multiplied by a Gaussian envelope of width $\sigma$:
\begin{equation}
    \psi_{\ell}^m(\mathbfit{u}) = \frac{\exp{\left[-|\mathbfit{u}|^2/(2\sigma^2)\right]}}{(2\pi\sigma^2)^{3/2}}|\mathbfit{u}|^\ell Y_\ell^m(\mathbfit{u}/|\mathbfit{u}|).
\end{equation}
The wavelets dilate to different scales as described by $j$:
\begin{equation}
    \psi_{j,\ell}^m(\mathbfit{u}) = 2^{-3j}\psi_{\ell}^m(2^{-j}\mathbfit{u}).
\end{equation}
We will fix $\sigma=1$ for the remainder of this work, which is also what~\cite{Eickenberg18} do in effect for their work.\footnote{This will also happen to sidestep potential issues with normalisation in implementation---note for example that the functional form of the mother wavelet as defined by~\cite{Valogiannis22b,Valogiannis22a} does not include division by $\sigma^3$ as we do, despite their use of $\sigma=0.8$.}

With the family of mother and dilated wavelets defined, we can explicitly write down the first- and second-order wavelet scattering coefficients $S_1$ and $S_2$ given an input field $I(\mathbfit{u})$ (again, as a function of image coordinates). We also introduce an extra parameter $q$, by which we exponentiate the complex modulus operation:
\begin{equation}
    S_1(j,\ell;q) = \avg{\left(\sum_{m=-\ell}^\ell|I(\mathbfit{u})*\psi_{j,\ell}^m|^2\right)^{q/2}}_\mathbfit{u}.\label{eq:S1_raw}
    \end{equation}
We only evaluate second-order coefficients across different scales, and not across different $\ell$. Then for scales $j$ and $j'>j$,
    \begin{equation}
    S_2(j,j',\ell;q) = \avg{\left(\sum_{m'}\left|\left(\sum_m|I(\mathbfit{u})*\psi_{j,\ell}^m|^2\right)^{1/2}*\psi_{j',\ell}^{m'}\right|^2\right)^{q/2}}_\mathbfit{u}.\label{eq:S2_raw}
\end{equation}
Since all coefficients involve sums in $m$, the calculation discards rotational information, but the variation with $\ell$ still encodes some level of angular or morphological information. Given that our wavelet basis is given by the solid harmonics, we should expect the variation with $\ell$ to specifically encode the different amounts of isotropic power contained in field monopoles, dipoles, quadrupoles, octopoles, hexadecapoles, and so on.\added{ 

With our choice of solid harmonic wavelets, only the isotropic power is encoded as orientation information is lost in the complex modulus operation, but a different choice of wavelet family could encode further information in the WST, which may be desirable given the expected redshift-space distortion of the CO signal. However, for this work we continue with the solid harmonic WST as it has a readily available implementation in Kymatio, and we leave exploration of oriented wavelets in the context of CO LIM to other future work.}

The choice of $q$ determines the relative weighting of overdensities against underdensities in determining the WST coefficients. Given the noisy nature of LIM data cubes, a value of $q>1$ to excite a moderate but not excessive emphasis on overdensities is likely desirable. We can calculate the WST with $q=\{0.5,0.75,1.0,1.5,2.0\}$ to examine detection significance, but beyond that we will consider only the optimal value of $q$ out of those five choices, rather than consider multiple values of $q$ at once. This is to avoid further worsening the conditioning of the covariance matrix, as we expect many of the WST coefficients to be strongly correlated across different values of $q$.

Note that~\cite{kymatio} describe a low-pass filter convolution at each step after the modulus operation. For the solid harmonic wavelet transform as implemented by Kymatio at time of writing, this is done with a simple summing operation (equivalent to integration over the volume as described in~\cite{Eickenberg18}, given $\mathbfit{u}$ is in image coordinates). We simply divide the output from Kymatio by the number of voxels to obtain the spatial average described above. The difference between averaging and summing is immaterial to the key results of this work like detection significance and parameter constraints.

As we have discussed previously, the Universe is \emph{mostly} homogeneous and therefore one might suppose that there should be greater information contained in variation of the coefficients with scale $j$ than in variation with multipole index $\ell$. %Any anisotropies are more likely to be buried under noise in a LIM observation in particular: for example, the Pathfinder phase of COMAP could detect the monopole power spectrum with total signal-to-noise of $\approx9$ yet only make a marginal $\sim3\sigma$ detection of the quadrupole power spectrum encapsulating the leading anisotropies. By analogy, averages of the WST coefficients over $\ell$ should contain the majority of the information content of the WST.
Given this expectation, we propose the following reduced three-dimensional WST coefficients that discard angular information:
\begin{align}
    s_1(j;q)&\equiv2^j\avg{S_1(j,\ell;q)}_{\ell\geq0};\label{eq:s1_sl}\\
    s_2(j,j';q)&\equiv2^{j'-j}\avg{S_2(j,j',\ell;q)/S_1(j,\ell;q)}_{\ell>0};\label{eq:s2_sl}\\
    s_{20}(j;q)&\equiv\avg{S_2(j,j',\ell=0;q)/S_1(j,\ell=0;q)}_{j'>j}.\label{eq:s20_sl}
\end{align}
The calculation of $s_1$ averages entirely over angular information, while the reduced second-order coefficients separate out $\ell=0$ as a degenerate case (since $Y_0^0$ is a constant) to obtain $s_{20}$ but then still average over $\ell>0$ to obtain $s_2$. This should provide information about peaks across different scales and interactions between scales, but discard all shape information. Therefore we will dub this the `shapeless' WST coefficient set.

We also consider a less aggressively reduced set of WST coefficients that does preserve variation with $\ell$ and therefore shape information as well as scale-dependence:
\begin{align}
    \tilde{S}_1(j,\ell;q)&\equiv 2^jN_1(j,\ell)S_1(j,\ell;q);\label{eq:S1_res}\\
    \tilde{S}_2(j,j',\ell;q)&\equiv2^{j'-j}N_2(j,j',\ell)S_2(j,j',\ell;q)/S_1(j,\ell;q).\label{eq:S2_res}
\end{align}
We excise $\tilde{S}_2(j,j',\ell=0)$ for all $j'>j+1$ from this coefficient set, due to the fact that these correlate almost perfectly with $\tilde{S}_2(j,j'=j+1,\ell=0)$ in practice. With that excision we have fully defined what we will refer to as the `rescaled' set (summarised with the raw and shapeless WST coefficients in~\autoref{tab:coeffdef}). Here $N_1$ and $N_2$ are arbitrary normalisation factors to be defined later that will not depend on $S_1$, $S_2$, or the input field in general. Rather $N_1$ and $N_2$ will be defined to scale the data in a way that optimises the condition number of the estimated covariance matrix, which we will discuss further in~\autoref{sec:cov}. Like the shapeless coefficients, the second-order coefficients are rescaled by appropriate first-order counterparts to significantly reduce correlations between coefficients, which is also key for covariance matrix conditioning.

The choice of these forms is broadly informed by the orientation reductions and normalisations discussed by~\cite{Cheng21}, but also by the correlations of the raw WST coefficients in simulations that we will show later. The prefactors of $2^j$ and $2^{j'-j}$, present in both the rescaled and shapeless coefficients, are immaterial to correlations but serve to normalise the coefficients against general trends of the coefficient values in presentation, which should improve conditioning of the covariance matrix. In particular, while we will need to define optimal $N_1$ and $N_2$ formulae for the full rescaled coefficients based on the behaviour of the signal and noise, the results of~\autoref{sec:cov} will show that the shapeless coefficients tend to result in well-conditioned covariance matrix estimates without further fine-tuning.

The shapeless coefficients also significantly reduce the dimensionality of any linear algebra involved. If $j$ ranges from 0 to $J$ and $\ell$ ranges from 0 to $L$ -- where for this work we will use $J=4$ and $L=3$ -- the number of raw WST coefficients for fixed $q$ is $(J+1)(J+2)(L+1)/2=60$, of which we only omit $J(J-1)/2=6$ when considering the rescaled coefficient set. Meanwhile, we have $J+1$ $s_1$ coefficients, $J$ $s_{20}$ coefficients (since $s_{20}(J;q)$ is not a valid calculation), and $J(J+1)/2$ $s_2$ coefficients for a total of $J(J+5)/2+1=19$ shapeless WST coefficients.

\begin{table}
    \centering
    \begin{tabular}{r|c}
         \hline Label & Definition \\\hline
         Raw & \parbox[c]{6.5cm}{\begin{itemize}\item $S_1(j,\ell;q)$ as defined in~\autoref{eq:S1_raw}, for all $j\in\{0,\dots,J\}$ and $\ell\in\{0,\dots,L\}$
         \item $S_2(j,j',\ell;q)$ as defined in~\autoref{eq:S2_raw}, for all $j\in\{0,\dots,J-1\}$, $j'>j$, and $\ell\in\{0,\dots,L\}$\end{itemize}}\\\hline
         Rescaled & \parbox[c]{6.5cm}{\begin{itemize}\item$\tilde{S}_1(j,\ell;q)$ as defined in~\autoref{eq:S1_res}, for all $j\in\{0,\dots,J\}$ and $\ell\in\{0,\dots,L\}$
         \item $\tilde{S}_2(j,j',\ell;q)\propto S_2(j,j',\ell)/S_1(j,\ell)$ as defined in~\autoref{eq:S2_res}, for all $j\in\{0,\dots,J-1\}$, $j'>j$, and $\ell\in\{0,\dots,L\}$ -- but omit if $j'>j+1$ and $\ell=0$\end{itemize}}\\\hline
         Shapeless & \parbox[c]{6.5cm}{\begin{itemize}\item$s_1(j;q)\propto\avg{S_1(j,\ell)}_\ell$ as defined in~\autoref{eq:s1_sl}, for all $j\in\{0,\dots,J\}$
         \item $s_2(j,j';q)\propto \avg{S_2(j,j',\ell)/S_1(j,\ell)}_{\ell>0}$ as defined in~\autoref{eq:s2_sl}, for all $j\in\{0,\dots,J-1\}$ and $j'>j$
         \item $s_{20}(j;q)\propto \avg{S_2(j,j',\ell=0)/S_1(j,\ell=0)}_{j'>j}$ as defined in~\autoref{eq:s20_sl}, for all $j\in\{0,\dots,J-1\}$\end{itemize}}
         \\\hline
    \end{tabular}
    \caption{Definitions of the different sets of coefficients derived from the WST, as used in this work.}
    \label{tab:coeffdef}
\end{table}

\section{Forecast methodology}
\label{sec:methods}
With our summary statistics defined, we move to outlining methods for modelling CO line emission (\autoref{sec:linemodel}), simulating COMAP observations (\autoref{sec:sims}), choosing bins for summary statistics (\autoref{sec:statbins}), and projecting parameter constraints in a Fisher matrix analysis (\autoref{sec:fisher}).
\subsection{CO line model}
\label{sec:linemodel}
\replaced{We use the CO(1--0) model of~\cite{COMAPESV}, which}{CO line emission in truth involves a variety of small-scale environmental and dynamical factors that are well beyond the scope of this paper\footnote{\added{For further details, we refer the interested reader to the review of~\cite{CW13}; the works of~\cite{linewidths} and~\cite{COMAPESV}, from which we take the form of our model; and references in each work. Again, the state of physical understanding and observational constraints around high-redshift CO line emission is unfortunately well beyond the scope of the present work.}}, whose focus is on recovery of large-scale non-Gaussian information. For our purposes, it is sufficient to use a simple halo model that} associates the virial mass $M_h$ and cosmological redshift $z$ of a dark matter halo with a CO luminosity $L$\added{. Specifically we use the empirical CO(1--0) model of~\cite{COMAPESV}, which assumes no redshift dependence and relates $M_h$ to $L$} via a double power law:
\begin{equation}\frac{L(M_h)}{L_\odot} = \frac{10^C}{(M_h/M)^A+(M_h/M)^B}.\label{eq:Lprime_of_M}\end{equation}
\replaced{N}{Here, $C$ represents an overall normalisation, $A$ and $B$ define the slopes of the double power law at extreme masses, and $M$ is a characteristic mass scale that defines the double power law turnabout point along with $C$. Compared to the presentation in~\cite{COMAPESV}, n}ote the slight tweak to how $C$ is defined in relation to the normalisation of the double power law, which is now $10^C$ rather than $C$, and against units of $L_\odot$ rather than K km s$^{-1}$ pc$^2$. The model also prescribes a mean-preserving log-normal scatter around the above relation, with the scatter width $\sigma$ expressed in units of dex.

Note that in the context of this model, we define $A$ and $B$ such that $A<B$, so that the $L(M_h)$ relation follows a power law with exponent $-A$ at low mass $M_h\ll M$ and a different power law with exponent $-B$ at high mass $M_h\gg M$.\added{ Overall, the double power law parameterisation of $L(M_h)$ reflects the expectation (considered in more detail in the original presentation of the model by~\citealt{COMAPESV}) that gas content grows and thus CO emission brightens with halo mass for low-mass objects, but feedback and quenching stunt this growth at sufficiently high mass.}

We adopt fiducial values based on the UM+COLDz model of~\cite{COMAPESV}, \replaced{such that}{which combines priors from the~\cite{Behroozi19} empirical model of the galaxy--halo connection and observational CO luminosity function constraints from~\cite{COLDzLF}. The parameter values thus chosen are} $A=-2.75$, $B=0.05$, $C=6.3$ (after accounting for the change in normalisation compared to~\citealt{COMAPESV}), $\log{(M/M_\odot)}=12.3$, and $\sigma=0.42$. This is also the fiducial model of~\cite{linewidths}.

To reflect the finite size of CO line profiles due to peculiar velocities within each dark matter halo, we also use the model of~\cite{linewidths} to associate a CO line width with each halo based on its virial velocity, and apply appropriate line broadening when generating CO signal cubes.\added{ We follow the implementation of this model by~\cite{COMAPESV}, and refer the reader to that work for further details. We also assert that small-scale effects outside of this line broadening (e.g., emission from satellite galaxies around each central dark matter halo) will be negligible in the context of a LIM survey, which probes large-scale line-intensity fluctuations rather than individual sources.}

\subsection{Simulations}
\label{sec:sims}

\begin{table}
    \centering
    \begin{tabular}{l|c|c}\hline
        Observational parameter & \multicolumn{2}{c}{COMAP phase:}\\& Pathfinder & COMAP-ERA \\\hline
        Frequency coverage (GHz) & 26--34 & 26--34\\
        Channel bandwidth (MHz) & 31.25 & 31.25\\
        Solid angle per field (deg$^2$) & 4 & 4\\
        Beam FWHM (arcmin) & 4.5 & 4.5\\
        Map pixel size (arcmin) for: &--&--\\
        \textbullet\ $P(k)$ and WST analyses & 2 & 2\\
        \textbullet\ VID analysis & 4 & 4\\
        Number of fields & 3 & 3\\
        Noise per WST input voxel (\textmu K) at:&--&--\\
        \textbullet\ 26--28 GHz & 21.2 & 4.5\\
        \textbullet\ 28--30 GHz & 18.7 & 4.0\\
        \textbullet\ 30--32 GHz & 18.2 & 3.9\\
        \textbullet\ 32--34 GHz & 19.8 & 4.2\\\hline
    \end{tabular}
    \caption{COMAP Ka-band survey parameters for the Pathfinder and COMAP-ERA phases, as used for the simulations in this work.}
    \label{tab:COMAPparams}
\end{table}

\autoref{tab:COMAPparams} defines survey parameters for two phases of Ka-band (26--34 GHz) observations with COMAP. The first `Pathfinder' phase is nominally a five-year campaign with the COMAP Pathfinder instrument, which has already been in mostly continuous operation for over three years at the time of writing. The second phase considered is the COMAP Expanded Reionisation Array (COMAP-ERA) concept outlined by~\cite{Breysse21}, which involves deployment of additional Ka-band feeds and extended survey operations.

In the context of this work, the only difference between the Pathfinder and COMAP-ERA phases is the expected noise per voxel, which have been extrapolated from noise levels in COMAP Field 1 observations rather than directly calculated from survey parameters.

\begin{itemize}
    \item For the Pathfinder experiment, we make the same assumption as~\cite{COMAPESV} that the noise per voxel will decrease relative to COMAP Early Science data by a factor of $\sqrt{40}$. The resulting Pathfinder projections broadly agree with the 17.8 \textmu K value quoted for simulations in~\cite{COMAPESV}.
    \item For COMAP-ERA, we use the assumption of~\cite{Breysse21} that COMAP accrues 1000 hours of integration time per field per year (implying that the COMAP Pathfinder survey should be equivalent to 5000 hours of integration time per field), and their projection that the COMAP-ERA survey should accrue the equivalent of 110000 Pathfinder hours. This implies a further improvement in noise per voxel by a factor of $\sqrt{22}$ when going from COMAP Pathfinder data to COMAP-ERA data. The assumed hours per field per year (and therefore the factor obtained here) is credible to within a factor of order unity, certainly a much smaller factor than the level of noise improvement being forecast between COMAP Early Science data and COMAP-ERA data.
    \end{itemize}
    
 \cite{Breysse21} also define an intermediate COMAP Epoch of Reionisation (COMAP-EoR) experiment, but for our purposes it is sufficient to consider the two extremes of near-term and far-future CO LIM datasets, with their corresponding noise levels.

We will want to generate large numbers of simulations of such datasets, as numerical estimation of WST covariances is necessary in the absence of their analytical understanding. Fortunately, the peak-patch method~\citep{Stein19} grants us dark matter halo catalogues of sufficient accuracy for thousands of simulated COMAP observations. In particular, we begin with the same set of peak-patch simulations used by~\cite{Ihle19}, which comprises 161 independent lightcone simulations across a comoving box of volume $L_\text{box}^3=(1140\,$Mpc$)^3$ with a grid resolution of $N_\text{cells}=4096^3$. As the lightcone extent is equivalent to $9.6^\circ\times9.6^\circ$ in transverse dimensions and 26--34 GHz in CO(1--0) observing frequency, we split the halo catalogue associated with each lightcone into 16 mock COMAP fields to simulate 2576 semi-independent COMAP observations at $z\sim3$.

We translate the halo catalogue in each field into a simulated CO signal using \texttt{limlam\_mocker}\footnote{\url{https://github.com/georgestein/limlam\_mocker}} -- including line broadening as mentioned in~\autoref{sec:linemodel} -- and apply angular smoothing by a Gaussian beam with full width at half maximum of 4.5 arcmin, matching the COMAP instrumental resolution. \added{We show an example realisation of the CO signal at this stage in~\autoref{fig:COviz}.

\begin{figure}
    \centering
    \includegraphics[width=0.96\linewidth,clip=True,trim=0 3mm 0 3mm]{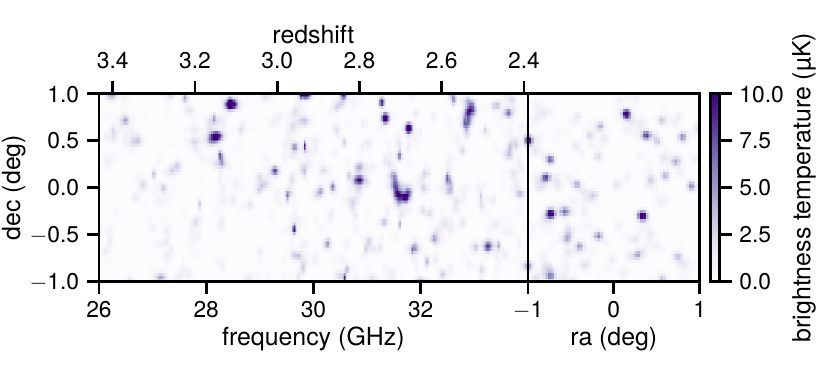}
    \caption{Example realisation of the CO signal, including line broadening and angular smoothing but excluding further high-pass filtering and additive Gaussian noise. We show the signal at two faces of the simulated volume.}
    \label{fig:COviz}
\end{figure}

}On top of these small-scale effects, we can appropriately account for the removal of large-scale modes from the signal expected from the COMAP pipeline. Appendix D of~\cite{COMAPESV} suggests an approximation for the COMAP power spectrum transfer function in early science data, with the following component in particular likely associated with high-pass filtering and subtraction of large-scale spectral modes:
\begin{equation}
    \mathcal{T}_\text{hp}(k_\perp,k_\parallel) = \frac{1}{(1+e^{5-100\text{\,Mpc}\cdot k_\perp})(1+e^{5-200\text{\,Mpc}\cdot k_\parallel})},
\end{equation}
with $k_\parallel$ denoting the magnitude of the line-of-sight component of the wavevector $\mathbfit{k}$ for which we evaluate the transfer function, and $k_\perp$ denoting the magnitude of the transverse part of the wavevector (such that $k_\perp^2+k_\parallel^2=k^2$).

We apply an equivalent Fourier filter to our simulated data cubes by taking the Fourier transform, multiplying each point of the Fourier-space data cube by $\mathcal{T}_\text{hp}^{1/2}$ (since $\mathcal{T}_\text{hp}$ is effectively the transfer function for the squared magnitude of the Fourier transform), and inverting the Fourier transform to obtain the high-pass filtered CO temperature cube. The final simulated COMAP observation includes uniform Gaussian noise added to this filtered signal, with noise per voxel as specified in~\autoref{tab:COMAPparams} for each COMAP phase.

We will consider each 2 GHz sub-band of the total frequency coverage separately, and in particular we will excise four frequency channels at each edge of each sub-band. This leaves a data cube spanning 60 voxels across each angular dimension -- coarsened to 30 voxels on each side for the VID analysis -- and 56 voxels along the line of sight. This results in a roughly isotropic cube in comoving space, corresponding to a $\approx220\times220\times240$ Mpc$^3$ cube at the central redshift of $z\approx2.8$.

\subsection{Bins for statistics}
\label{sec:statbins}
We have already significantly detailed the power spectrum, the VID, and the WST in~\autoref{sec:stats}. Here we briefly clarify the scale or temperature bins used for each of these statistics.

As stated above, for $P(k)$ and WST analyses we use a temperature cube of $60\times60\times56$ voxels for each 2 GHz sub-band, with each voxel spanning roughly $3.6\times3.6\times4.2$ Mpc$^3$. The extent and grid size of the cube suggests $k_\text{min}\sim0.03$\,Mpc$^{-1}$ and $k_\text{max}\sim0.8$\,Mpc$^{-1}$. This is reflected in the $k$-bins used for the power spectrum calculation, which are 29 linearly spaced bins with $\Delta k\approx0.03$\,Mpc$^{-1}$ (where the exact value depends on the sub-band) and the lower edge of the first bin (which contains as little as one Fourier mode) placed at $k=0$.

We use WST parameters previously outlined in~\autoref{sec:wst}, including a maximum scale index of $J=4$ and a maximum spherical harmonic multipole index of $L=3$. Each wavelet spans a band of Fourier wavenumbers, but if we associate a scale index $j$ roughly with a span of $2^j$ voxels (meaning it is essentially senseless to set $J$ above the base-2 logarithm of the extent of the data cube), the scales of $j=\{0,1,2,3,4\}$ correspond roughly to $\{4,8,16,32,64\}$ Mpc scales. If this is roughly the half-wavelength of a Fourier mode, the corresponding wavenumbers would be $k\sim\{0.8,0.4,0.2,0.1,0.05\}$ Mpc$^{-1}$. Therefore, the range of scales being probed is quite similar between $P(k)$ and the WST.

To better match the angular size of the CO map voxels to the COMAP beam width, we use a coarser pixel size for VID calculations, meaning we work with a cube of $30\times30\times56$ voxels. We generate voxel temperature histograms using 12 log-spaced bins whose edges span $10^1$--$10^{1.6}$\,\textmu K (or approximately 10--40\,\textmu K, limiting ourselves to probing $>4\sigma$ excesses above COMAP-ERA noise). This is likely a conservative choice, given that~\cite{Ihle19} used bins going up to 100\,\textmu K, but justified given our somewhat dimmer CO emission model and the lower noise level of COMAP-ERA simulations.
\subsection{Fisher formalism}
\label{sec:fisher}
We undertake an analysis in the Fisher matrix formalism\footnote{For more on the topic of Fisher analyses, Section 11.4 of~\cite{Dodelson} for a pedagogical overview and~\cite{Coe2009} for additional guidance in practice.} to consider potential constraints on the CO emission model outlined in~\autoref{sec:linemodel}. Given the necessary but tenuous assumption of Gaussian covariances, these should not be taken as quantitative forecasts, but qualitative comparisons of constraining power for different summary statistics should still be of interest.

For this work, we fix $B$ and $M$ at their fiducial values. In the basis of the remaining parameters $\{\lambda_r\}=\{A,{C},\sigma\}$, the Fisher matrix for an observable vector $\mathbfit{O}_\alpha$ with covariance matrix $\mathbfss{C}_{\alpha\beta}$ is given by
\begin{equation}
    \mathbfss{F}_{rs} = \sum_{\alpha,\beta}\frac{\partial{\mathbfit{O}}_{\alpha}}{\partial\lambda_r}\mathbfss{C}^{-1}_{\alpha\beta}\frac{\partial{\mathbfit{O}}_{\beta}}{\partial\lambda_s},\label{eq:fisher}
\end{equation}
approximating the derivatives with central difference quotients based on median values for the observables across 2576 simulated COMAP survey volumes (as described in~\autoref{sec:sims}) obtained at parameter values 0.05 above and below the fiducial value for each parameter. We impose no additional priors.

In fact, we have observable vectors for each one of four COMAP sub-bands, which we will index with $i$. To minimise effects of spurious artefacts in numerically estimated covariances, we evaluate $\mathbfss{F}_{i;rs}$ separately for each sideband and obtain a total Fisher matrix $\sum_i\mathbfss{F}_{i;rs}$ by simple summation. We thus treat the four COMAP bands as independent observations, which seems justifiable in practice as correlation coefficient matrices for all observables show negligible correlation ($\lesssim1\%$) across frequency sub-bands. Given the large comoving length spanned by the COMAP frequency coverage, this is perhaps not surprising, certainly for statistics at smaller scales.

Recall also that COMAP observes three independent survey fields, which should decrease covariances by a factor of three. We equivalently multiply the Fisher matrix by 3 to reflect this fact, then invert this Fisher matrix to obtain the parameter covariance matrix for the full COMAP Pathfinder or COMAP-ERA survey.

\section{Results}
\label{sec:results}

This section will eventually answer the questions we posed at the outset of this paper about the WST in the context of COMAP. However, it will be instructive to first qualitatively examine how the WST changes with our CO model parameters, which we do in~\autoref{sec:qualcomp}. We must also examine the covariance of the WST coefficients in~\autoref{sec:cov} before considering detection significances in~\autoref{sec:snr}, and both will play a role in our choice of $q$ for the WST for finally forecasting parameter constraints in~\autoref{sec:fisherres}.
\subsection{Qualitative examination of WST coefficients in CO line-intensity simulations}
\label{sec:qualcomp}
A key motivation for using the WST is the fact that the coefficients summarise scale interactions in an interpretable fashion, characterising not only total power at different scales but also information about morphology and sparsity. To demonstrate this, we ran an additional set of simulations for each one of our 2576 mock COMAP volumes with parameter values shifted $\pm0.1$ away from fiducial values. These are separate from the simulations used for our Fisher analyses, which shift parameter values only by $\pm0.05$. The larger shift is to better illustrate how the WST coefficients change in a qualitative comparison, although we can (and successfully do) use these simulations to also verify that our estimates of $\partial\mathbfit{O}_\alpha/\partial\lambda_r$ are relatively stable.

\begin{figure*}
    \centering
    \includegraphics[width=0.975\linewidth]{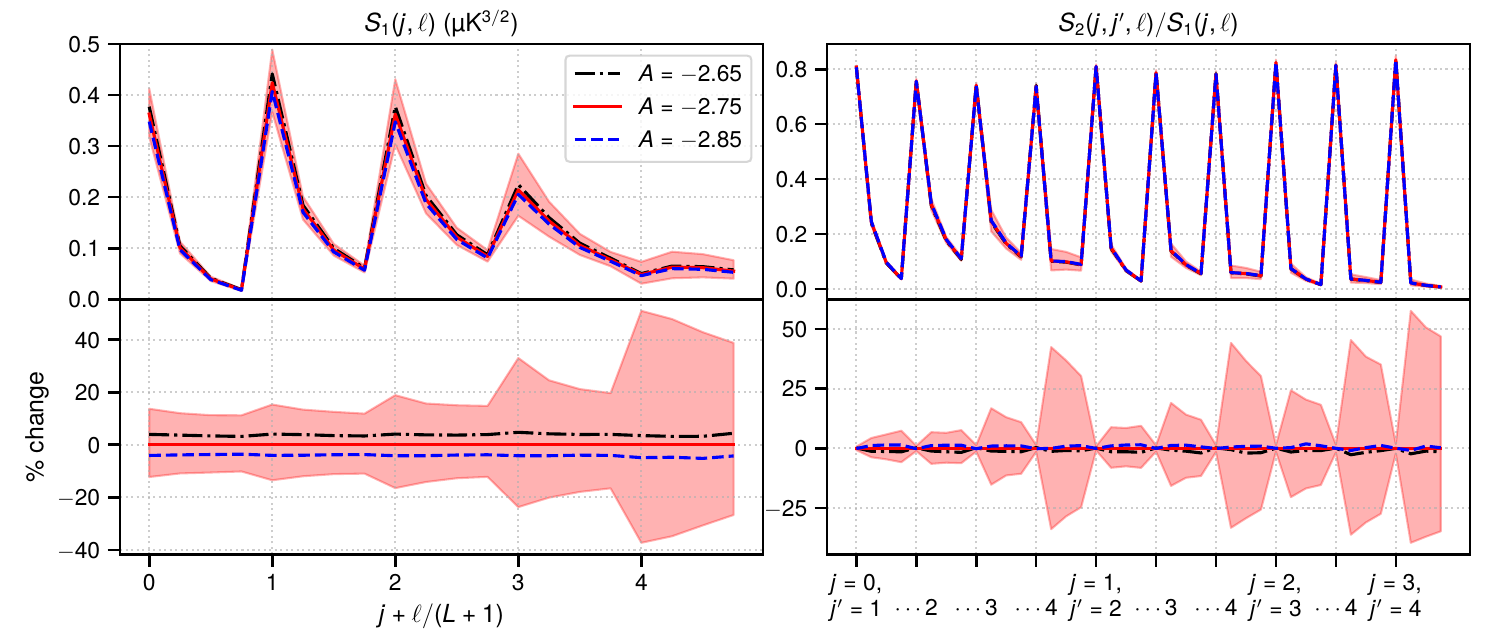}
    
    \includegraphics[width=0.975\linewidth]{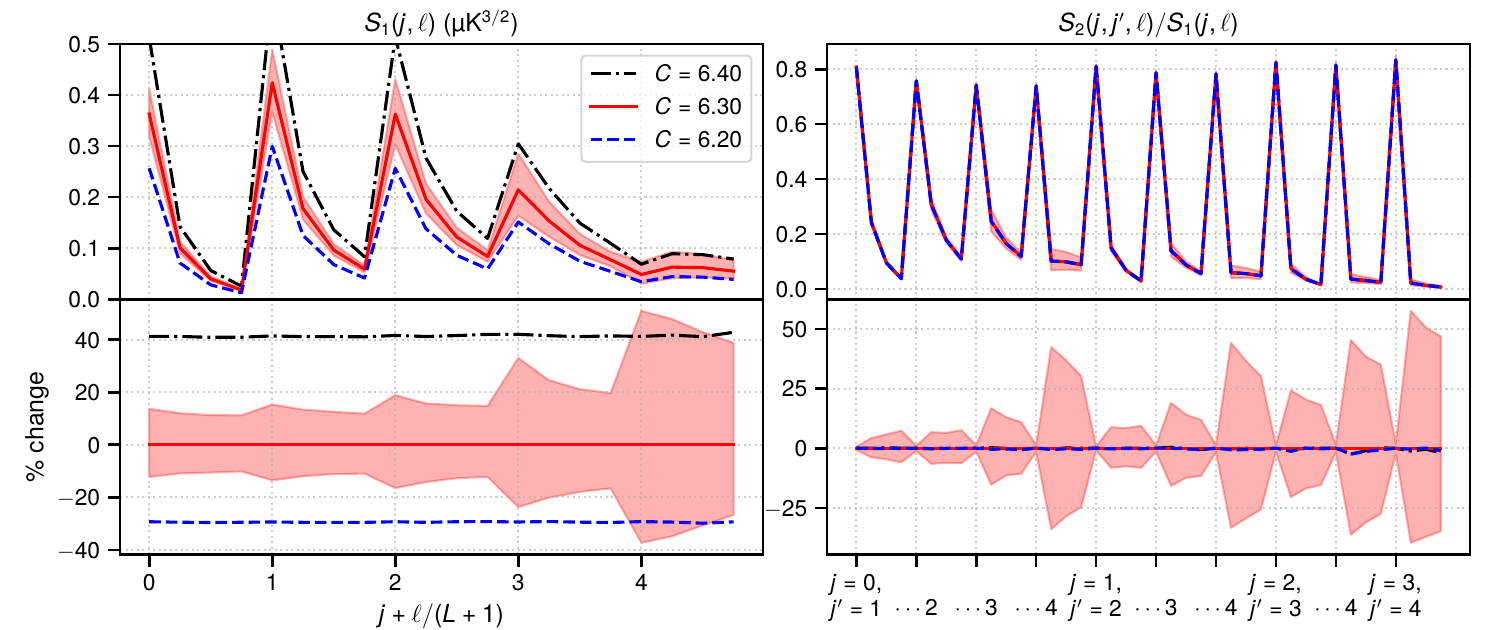}
    
    \includegraphics[width=0.975\linewidth]{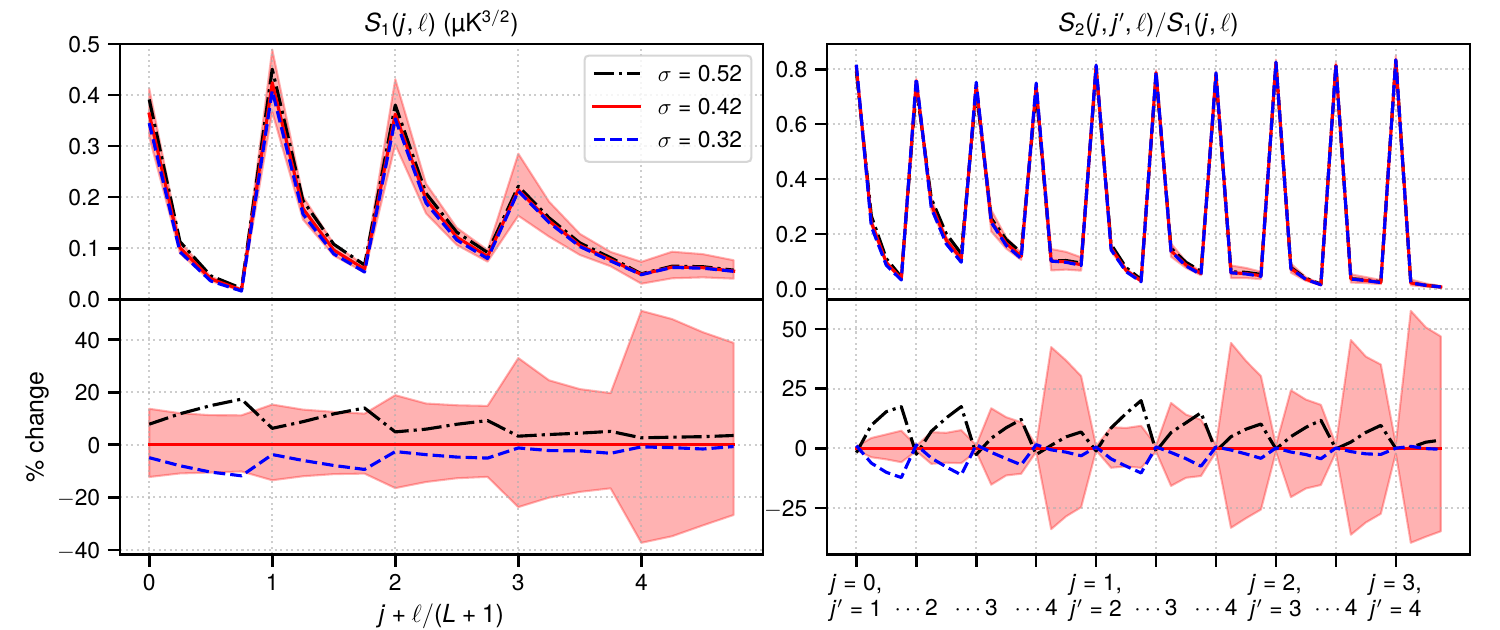}
    \caption{Variation of the median raw first-order WST coefficients $S_1(j,\ell;q=1.5)$ (left panels) and the partially rescaled second-order coefficients $S_2(j,j',\ell;q=1.5)/S_1(j,\ell;q=1.5)$ (right panels), given noiseless CO line-intensity simulations with changes in model parameters $A$ (upper panel sets), $C$ (middle panel sets) and $\sigma$ (lower panel sets). We show both the original values (upper panel in each panel set) and the change relative to the results for fiducial parameter values (lower panel in each panel set), as well as 68\% sample intervals (red shaded areas) around the median given the fiducial parameter values.}
    \label{fig:derivs_noiseless}
\end{figure*}

\autoref{fig:derivs_noiseless} shows WST coefficients for simulations with beam smoothing and line broadening but without the high-pass transfer function $\mathcal{T}_\mathrm{hp}$ or random Gaussian noise per voxel as described in~\autoref{sec:sims}. We specifically show the first-order WST coefficients $S_1(j,\ell;q=1.5)$ and the second-order WST coefficients by their first-order counterparts $S_2(j,j',\ell;q=1.5)/S_1(j,\ell;q=1.5)$, such that the latter are partially rescaled (noting that what we have termed the `rescaled' $\tilde{S}_2$ differs by additional factors of $2^{j'-j}$ and the as-yet-undetermined $N_2$).

We observe a few interesting general trends for the WST coefficients even considering their values for the fiducial parameter values alone. First, note that both $S_1(j,\ell)$ and $S_2(j,j',\ell)$ tend to decrease with higher multipole index $\ell$. Thinking about the variation of these coefficients with $\ell$ as encoding how much brightness is in different kinds of shapes that fill out the cosmic web, this suggests more brightness in simpler shapes -- i.e., the majority of observed CO luminosity comes from CO brightness `monopoles' rather than `dipoles', with more brightness contained in dipoles than `quadrupoles', and so on.

Note also the amount of relative variation in the WST coefficients, which appears to trend upwards with higher values of $j$, i.e., larger scales. This is not too surprising, and is at least qualitatively analogous to how the relative variation in the power spectrum $\sigma[P(k)]/P(k)$ scales simply as the inverse of the square root of the number of modes $\sim k^{-1}$, due to there simply being fewer large-scale modes to sample in a finite volume.

Most interesting, however, is how the first- and second-order coefficients respond differently to changes in different parameters. Changing $C$, the overall normalisation of $L(M_h)$, affects $S_1$ uniformly across all $j$ and $\ell$ but essentially has zero effect on $S_2/S_1$. Meanwhile, shifts in the other two parameters affect both $S_1$ and $S_2/S_1$. The first-order coefficients $S_1$ are somewhat uniformly sensitive to changes in $A$ -- which we recall is the negative of the faint-end $L(M_h)$ power law exponent -- and the reduced second-order coefficients $S_2/S_1$ are less sensitive. Meanwhile, changes in the log-normal scatter $\sigma$ around the mean $L(M_h)$ relation result in a strongly scale- and $\ell$-dependent change in $S_1(j,\ell)$, and significant changes in $S_2/S_1$.

All of this highlights the simultaneously rich and interpretable information that the WST coefficients contain. Since $C$ is an overall normalisation, shifting its value does not affect different scales or shapes of CO emission in any non-uniform way. Therefore, only $S_1$ is sensitive to it, while sparsity as measured by $S_2/S_1$ is unaffected. Meanwhile, shifting $A$ down will decrease (and shifting it up will increase) the contribution of faint CO emitters to the total cosmological signal, which makes the signal both dimmer overall and sparser (or brighter and less sparse when shifting $A$ up). This leads to both $S_1$ and $S_2/S_1$ changing, although perhaps with $S_2/S_1$ less sensitive.

Finally, shifting $\sigma$ up increases the scatter in the halo mass--luminosity relation, leading potentially to significantly brighter but also more sparsely distributed emitters. In the power spectrum, this effect manifests as an increase in the shot noise contribution to the total $P(k)$, which chiefly affects higher $k$ or smaller scales where clustering becomes subdominant. We see the same thing here with the WST coefficients, which are more sensitive to changes in $\sigma$ for lower $j$ and smaller scales. But beyond shifts in $S_1$, we also see shifts in $S_2/S_1$, demonstrating the effect that the change in $\sigma$ has not merely on the typical scale of CO fluctuations but on their morphology and sparsity. We also see increased sensitivity to $\sigma$ with higher $\ell$, apparently reflecting how increased scatter results in more luminosity being contained in more complex shapes.

\begin{figure*}
    \centering
    \includegraphics[width=0.98\linewidth]{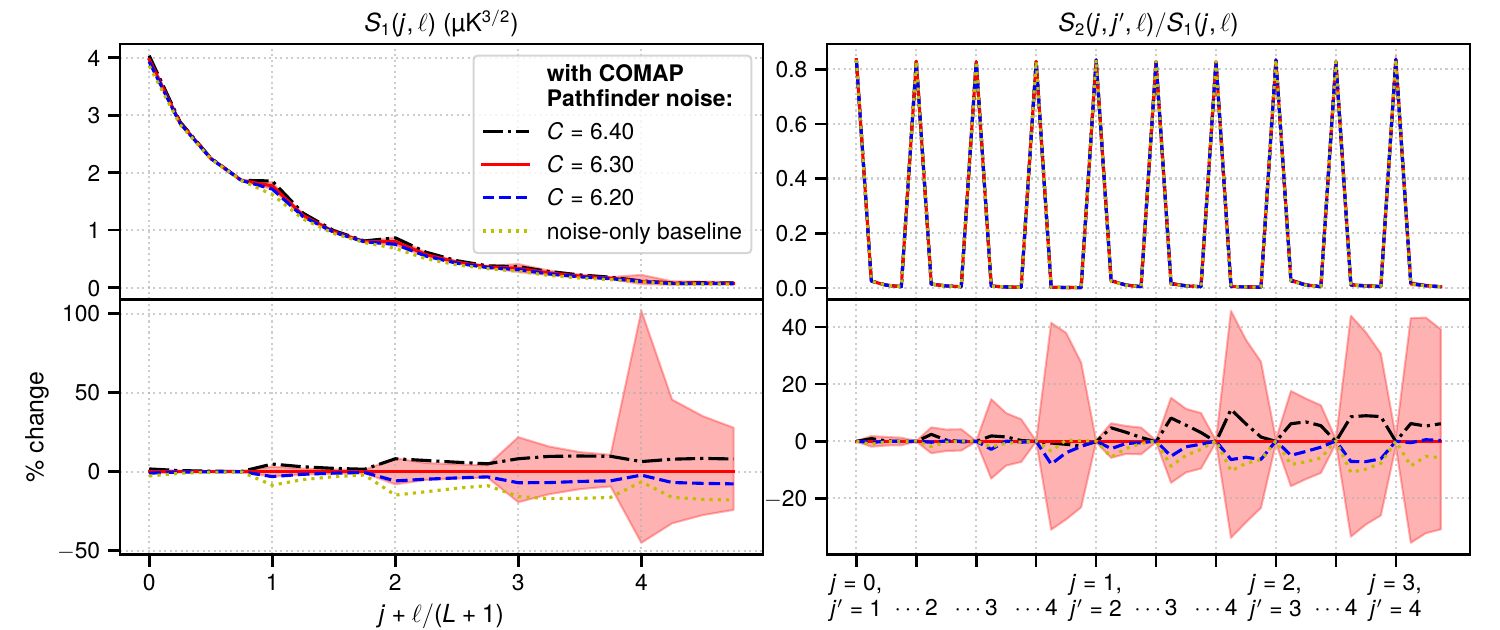}
    
    \includegraphics[width=0.98\linewidth]{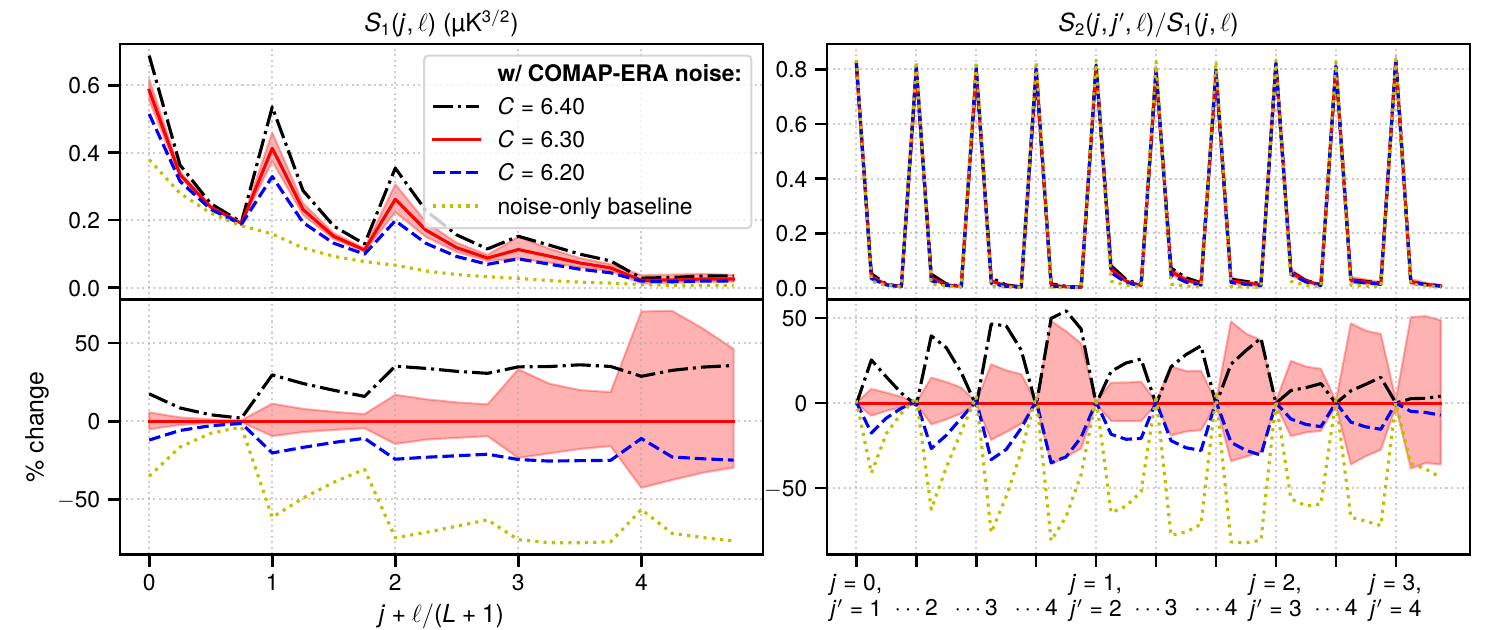}
    \caption{Same as~\autoref{fig:derivs_noiseless}, but only for the model parameter $C$, and given noisy CO line-intensity simulations with $\mathcal{T}_\mathrm{hp}$ applied to the CO signal as described in~\autoref{sec:sims}. We show results for both COMAP Pathfinder simulations (upper panel sets) and COMAP-ERA simulations (lower panel sets). We show a noise-only baseline (dotted) in addition to WST coefficients in the presence of signal with different values of $C$.}
    \label{fig:derivs_noisy}
\end{figure*}

All of this, however, is in the absence of observational effects. We must now consider the fact that real-world observations of CO line intensity will always be noisy, as well as being subject to various observational transfer functions like the $\mathcal{T}_\mathrm{hp}$ that we described in~\autoref{sec:sims}. The WST of a noisy data cube is very different from that of a noiseless data cube, because spaces that would have been line-intensity voids will be filled with spurious noise fluctuations, and features that would have been line-intensity peaks are effectively smeared out in a WST by the background noise. This is in fact what we see with noisy simulations in~\autoref{fig:derivs_noisy}. Where before noise only the first-order coefficients were sensitive to changes in $C$, now the reduced second-order coefficients $S_2/S_1$ are also sensitive. This makes sense because having brighter (or dimmer) CO contrast versus the noise level will in fact increase (or decrease) the sparsity of the observed temperature field.

Furthermore, in the case of COMAP Pathfinder simulations, the difference in $S_2/S_1$ from adding signal on top of noise is negligible compared to the variance due to the overall damping of sparsity in the observed temperature field. Therefore, in a first-generation CO line-intensity survey, only the first-order coefficients are likely to allow inferences about the CO field.

Decreasing the noise level from that expected for the COMAP Pathfinder survey to that expected for COMAP-ERA, we observe a corresponding decrease in the noise-only baseline (apparently by a factor of $22^{q/2}$, which we might expect from the factor-of-$22^{1/2}$ decrease in the noise level), and potentially detectable deviations in $S_2/S_1$ from the noise-only baseline. Note, however, that the interaction of noise and signal is not simply additive, even for first-order coefficients (ostensibly most analogous to $P(k)$). Comparing the first-order coefficients in~\autoref{fig:derivs_noiseless} to those in the COMAP-ERA case in~\autoref{fig:derivs_noisy}, the noise seems to add to $S_1$ at smaller scales but then drag down the CO signal to the noise-only baseline, with the effect being more marked on larger scales. This is also what~\cite{Greig22} appear to see for their two-dimensional $S_1$ coefficients at scales larger than the instrumental resolution. Given the nonlinear, multi-scale nature of the WST, we should not expect the interaction of signal and noise to be as simple as in a $P(k)$ analysis, and ultimately differentiating signal from noise in a WST analysis may require an even stronger understanding of both.

\subsection{Covariance of full and shapeless WST coefficient sets}
\label{sec:cov}

Reducing strong correlations between the components of our observable vector is not merely an aesthetic matter. Undertaking likelihood analyses and Fisher forecasts requires credible computation of the inverse of the covariance matrix $\mathbfss{C}$, as one may observe from~\autoref{eq:fisher}. High multicollinearity in the covariance matrix or other pathological behaviour will lead to the matrix inversion being an ill-conditioned problem, as measured by the condition number \replaced{$||\mathbfss{C}||\,||\mathbfss{C}^{-1}||$}{$\kappa\equiv||\mathbfss{C}||\,||\mathbfss{C}^{-1}||$} of the matrix (the norm of the matrix\added{\footnote{\added{The \texttt{numpy} implementation and thus this work both use the 2-norm in this context. With the 2-norm, the condition number is simply the ratio of the largest singular value to the smallest singular value. For a symmetric positive definite matrix, this is simply the ratio of the largest eigenvalue to the smallest eigenvalue. For further insights we refer the reader to the text of~\cite{GolubVanLoan96}.}}} times the norm of the inverse). We consider in~\autoref{tab:covcond} representative condition numbers for the full and shapeless WST coefficients across all values of $q$ initially considered.

\begin{table}
    \centering
    \begin{tabular}{cccc}
    \hline
         $q$ & \multicolumn{3}{c}{$\log_2{(||\mathbfss{C}||\,||\mathbfss{C}^{-1}||)}$ for covariance of:}\\
         & raw WST & shapeless WST & $s_1$ and $s_2$ only \\\hline
         $0.5$ & 28.4 & 19.5 & 16.0\\
         $0.75$ & 28.2 & 19.2 & 15.7 \\
         $1.0$ & 74.4 & 45.6 & 15.8 \\
         $1.5$ & 27.9 & 17.1 & 17.0 \\
         $2.0$ & 38.8 & 22.8 & 22.8 \\\hline
    \end{tabular}
    \caption{Base-2 logarithms of condition numbers of WST covariance matrices (which should be as low as possible for credible invertibility) estimated from the COMAP Pathfinder simulations in the 30--32 GHz frequency band including both signal and noise, for different reductions of the WST coefficients (or lack thereof). We do not show the condition numbers for rescaled WST coefficients as we only define them properly for $q=1.5$, but in that case we find the corresponding value is $\log_2{(||\mathbfss{C}||\,||\mathbfss{C}^{-1}||)}=18.0$.}
    \label{tab:covcond}
\end{table}

Ideally the condition number is as close to one as possible, as effectively it represents the factor by which inversion (or solutions of other linear algebra problems involving this matrix in general) can result in loss of precision. 32-bit and 64-bit floats respectively have 24 and 53 bits of total significand precision, so the base-2 logarithm of the condition number should not exceed these numbers. Most modern CPUs and programming languages use 64-bit float arithmetic by default, but in many cases GPU programming conventions (as might be used in GPU-accelerated WST calculations) still favour 32-bit floats for throughput and performance. With Kymatio at time of writing, even the CPU-based WST implementation uses a 64-bit complex float data type (i.e., 32-bit floats for each part of the complex number) in calculating the spatial `integral' (or sum), so the WST coefficients are encoded as 32-bit floats. This work will leave this choice unmodified, partly to make conservative use of memory and disk space, but partly because large numbers of WST evaluations will likely be necessary in future work and these may use GPU-accelerated calculations based on 32-bit float arithmetic. Therefore, WST coefficient sets that result in covariance matrices with a condition number above $2^{24}$ (or even within a factor of 2--3 of this threshold) should be treated with extreme suspicion.

With $q=1$ the raw WST covariance is absolutely unacceptable for inversion. \added{It is unlikely that the fact of the condition number exceeding our threshold is a fluke or numerical artefact, given the large number of realisations used to generate the covariance matrix, but we must be cautious. As an additional check, we show in~\autoref{fig:covcond_extra} covariance matrix condition numbers for additional values of $q$ close to 1. The results demonstrate that as $q$ approaches 1 from either direction, the condition number consistently ends up exceeding the acceptable range of values. That said, the condition number improves to an acceptable value when we discard information for $\ell=0$ (e.g., by discarding $s_{20}(j)$ from the shapeless WST coefficients). This suggests that as $q\to1$, the $\ell=0$ coefficients (which we recall correspond to a degenerate wavelet with constant amplitude) lead to some combination of extreme ranges of covariance values and pathological correlations.

\begin{figure}
    \centering
    \includegraphics[width=0.96\linewidth]{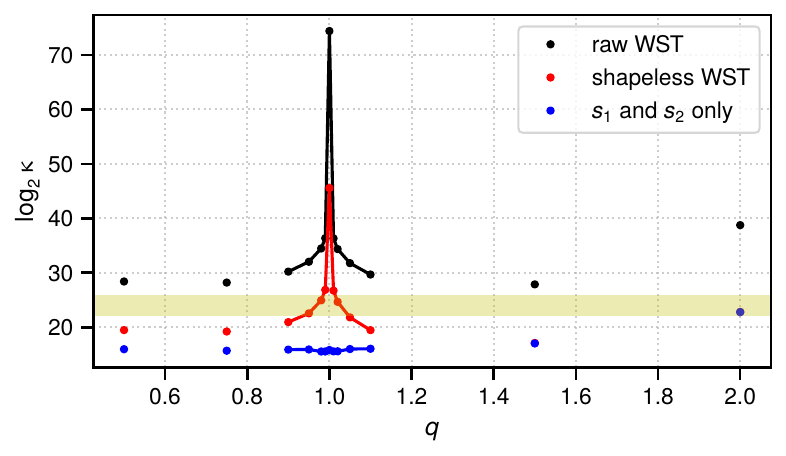}
    \caption{Plot of the same condition number logarithms given in~\autoref{tab:covcond} as a function of $q$, with additional calculations for extra values of $q$ that approach 1 from either direction. The condition numbers show a clear trend of rising to unacceptably high values as $q$ approaches 1 from either direction, unless we discard $\ell=0$ WST coefficients. We also show the approximate borderline of $\kappa\sim2^{24}$ (thick yellow band), above which inversion of the covariance matrix with 32-bit float arithmetic becomes unreliable, either because the matrix is ill-conditioned or because the matrix is outright singular.}
    \label{fig:covcond_extra}
\end{figure}

}With other values of $q$ inversion\added{ of the raw WST covariance} would be dangerous with 32-bit float arithmetic. By using the $\ell$-averaged WST coefficients (although \added{again, }in the case of $q=1$ we must discard $s_{20}(j)$ altogether), the condition number of the covariance matrix decreases by around a factor of $\sim10^3$, sufficient to bring it within the precision of 32-bit floats (albeit only by the equivalent of a couple of decimal digits). In the case of $q=2$, however, the condition number remains at an unacceptably high level.

\begin{figure*}
    \centering
    \includegraphics[width=0.48\linewidth]{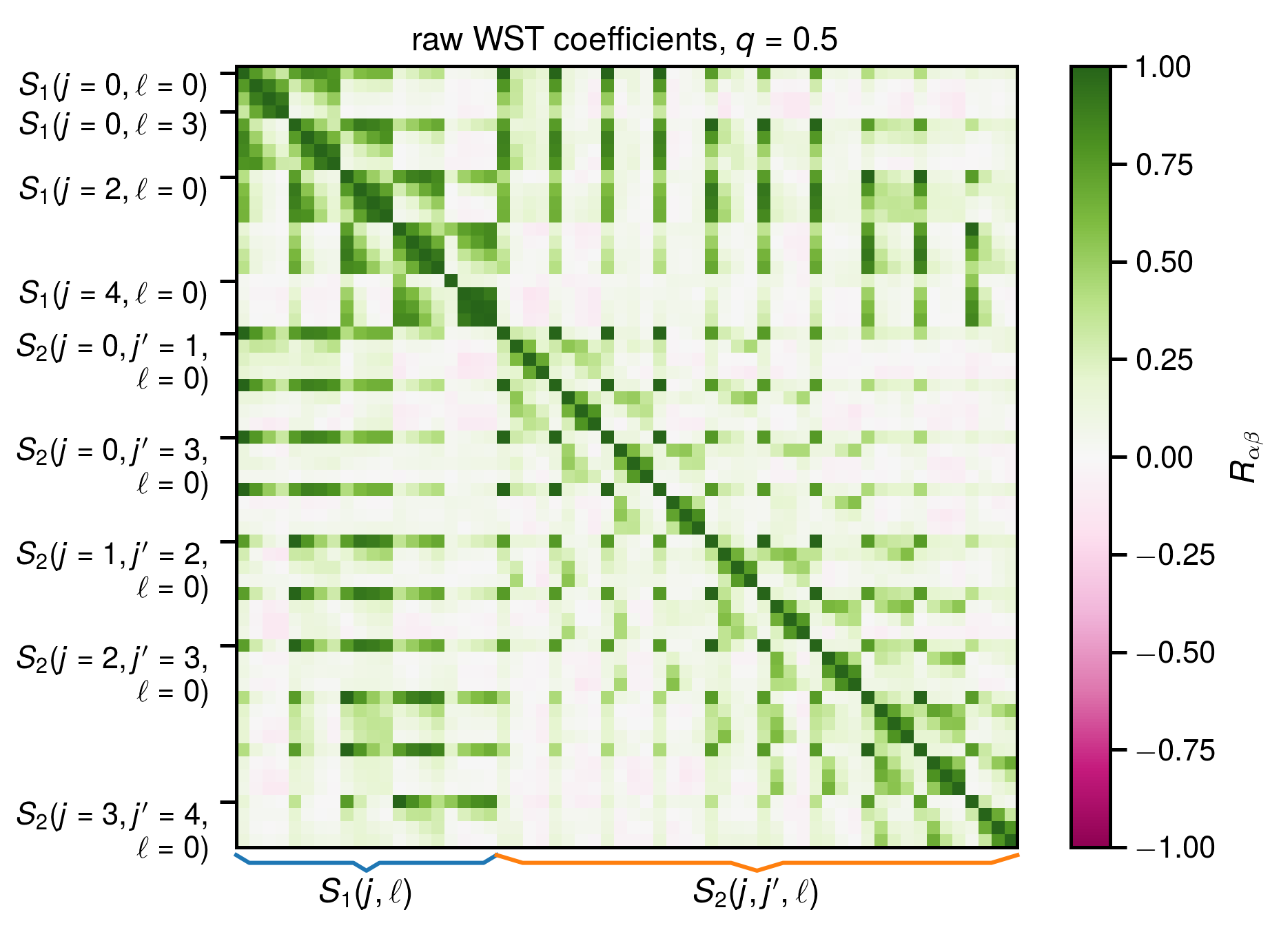}\quad\includegraphics[width=0.48\linewidth]{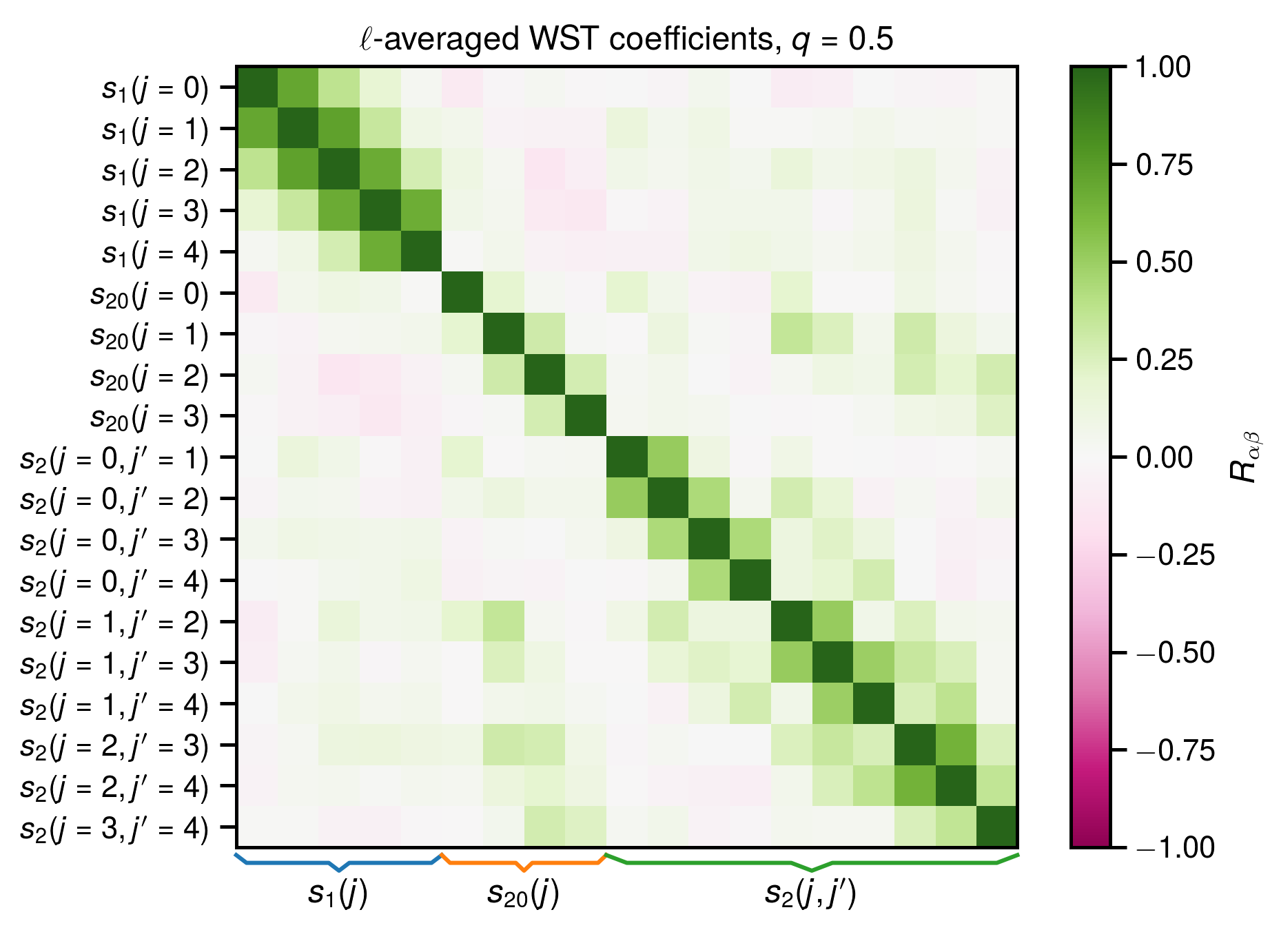}
    
    \includegraphics[width=0.48\linewidth]{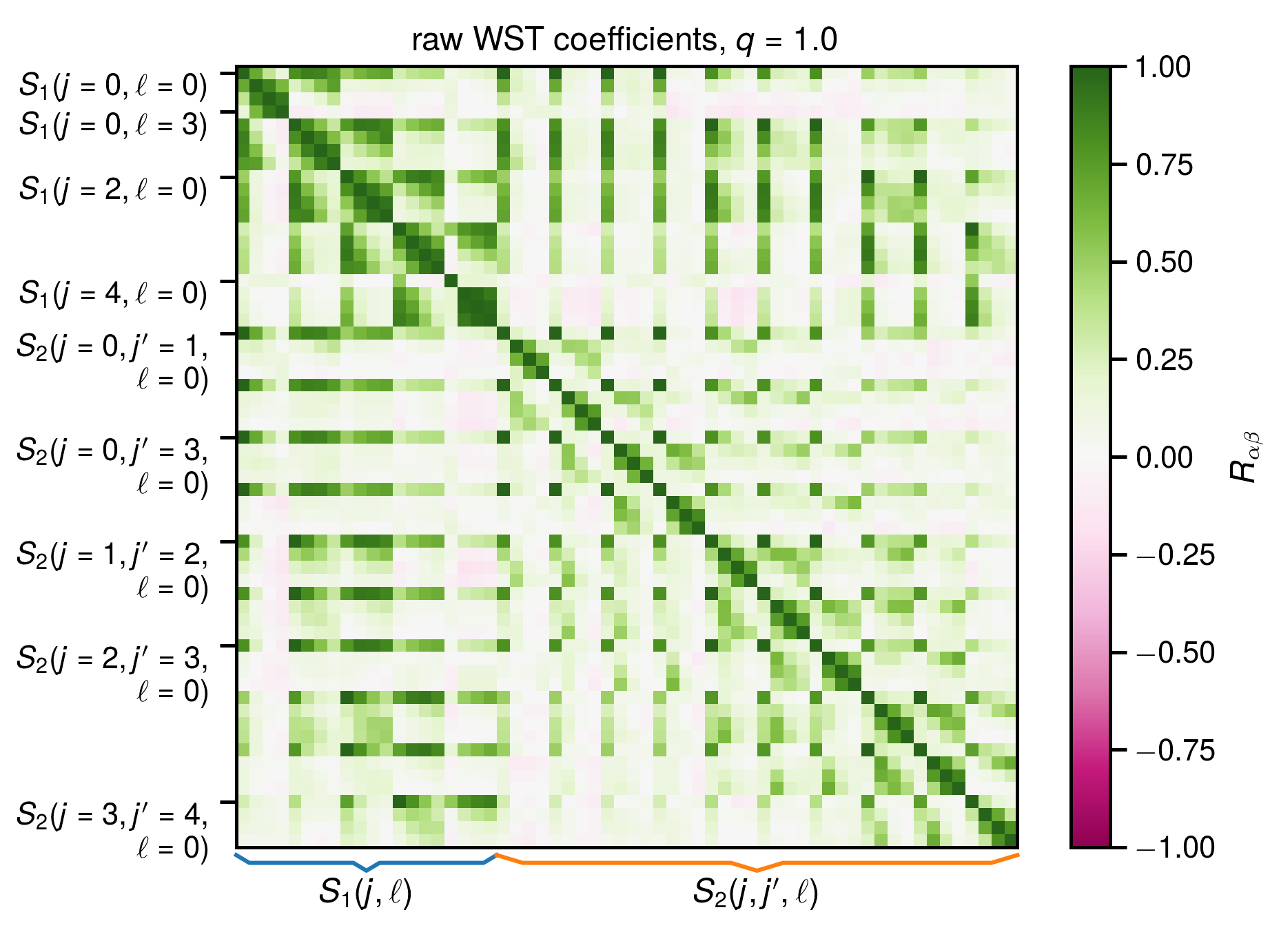}\quad\includegraphics[width=0.48\linewidth]{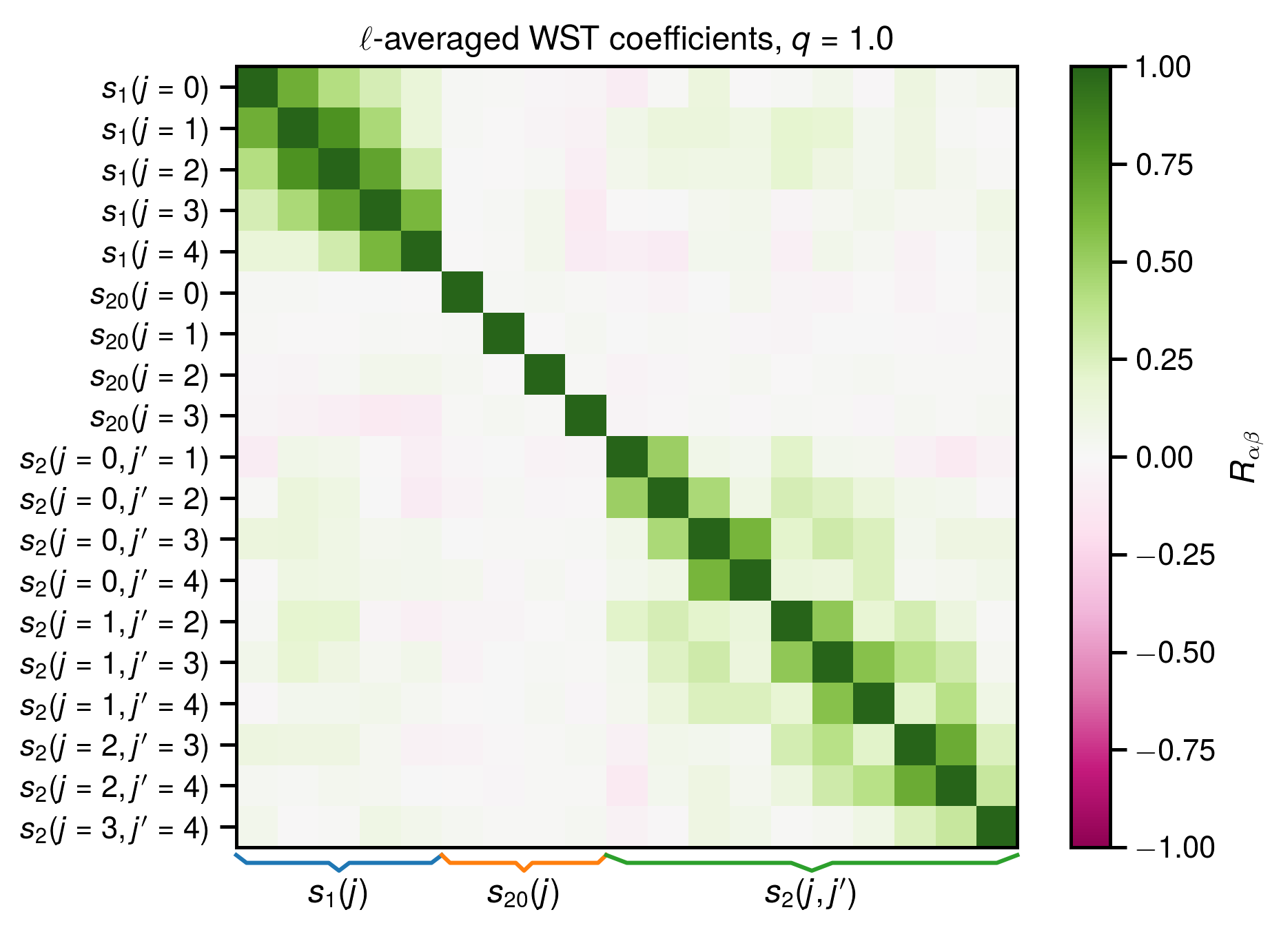}
    
    \includegraphics[width=0.48\linewidth]{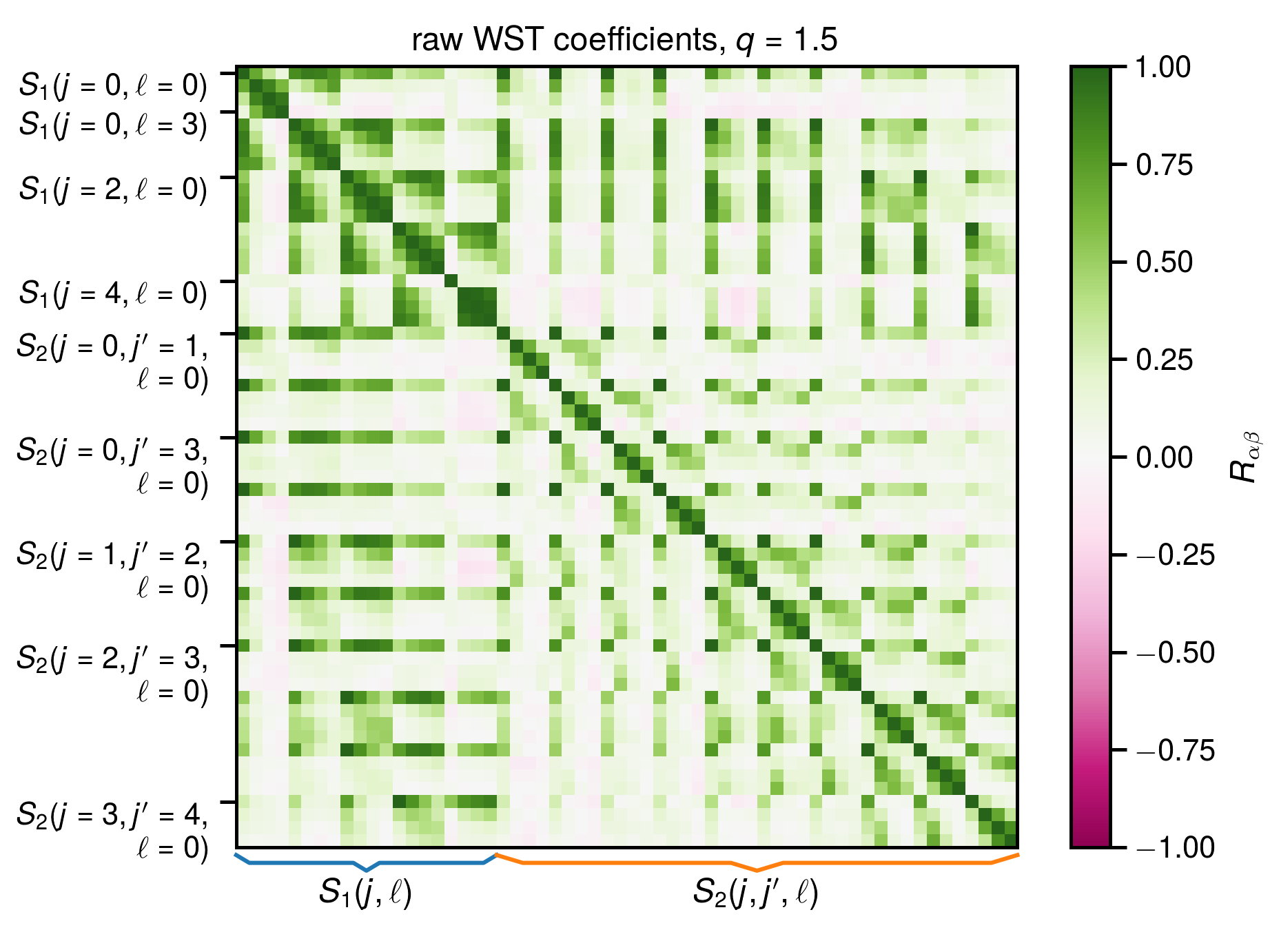}\quad\includegraphics[width=0.48\linewidth]{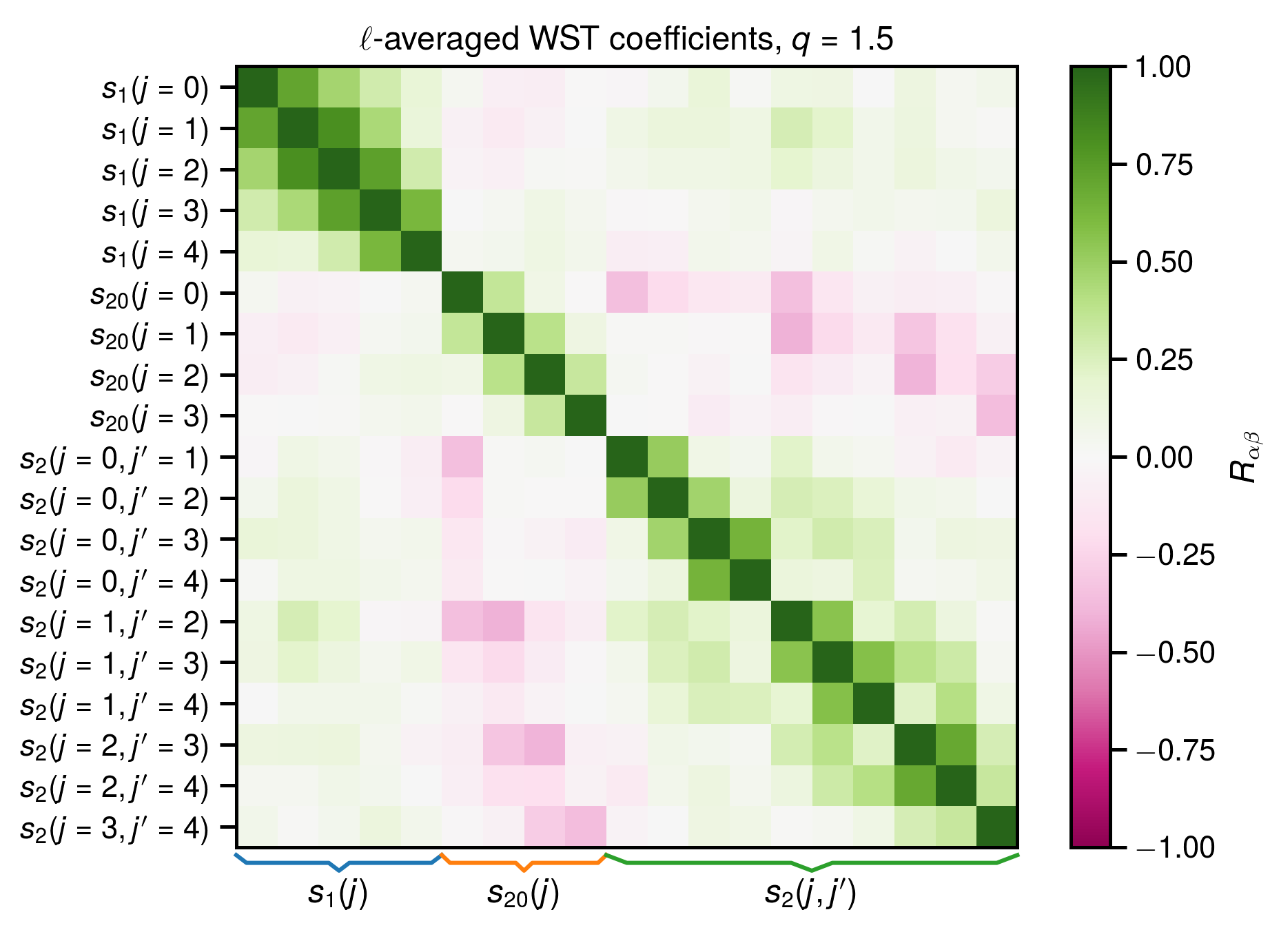}
    \caption{Correlation coefficient matrices $R_{\alpha\beta}$ for the raw WST coefficients (left panels) and the shapeless ($\ell$-averaged) WST coefficients (right panels) derived from the COMAP Pathfinder simulations in the 30--32 GHz band including both signal and noise, for three different values of the density-weighting exponent $q$ (top to bottom, indicated in annotations above each panel).}
    \label{fig:corr}
\end{figure*}

The high condition numbers of the covariance matrices $\mathbfss{C}_{\alpha\beta}$ are not entirely due to correlations alone, with the range of covariance values playing a large part also\added{\footnote{\added{This is clear when recalling the statement from the previous footnote that with the 2-norm, the condition number is simply the ratio of the covariance matrix's largest and smallest singular values (or eigenvalues if the covariance is positive definite).}}}. The latter effect can be discarded by examining the condition numbers of the correlation coefficient matrices $\mathbfss{R}_{\alpha\beta} = \mathbfss{C}_{\alpha\beta}/\sqrt{\mathbfss{C}_{\alpha\alpha}\mathbfss{C}_{\beta\beta}}$, and are what the covariance matrices would be if all variables were normalised by their standard deviation, We show correlation coefficient matrices in~\autoref{fig:corr} for selected $q$ for both raw and shapeless WST coefficients. The condition numbers for these matrices are significantly lower than the covariance matrix condition numbers, but in the case of $q=2$ the condition number of the correlation coefficient matrix for the full \replaced{raw (rescaled)}{raw} WST coefficient set is still in excess of \replaced{$2^{30}$ ($2^{29}$)}{$2^{30}$}. Overall, $q=2$ appears to result in highly pathological correlations between the WST coefficients \added{(not as easily understood as in the case of $q\to1$, and potentially requiring careful rescaling of coefficients beyond the scope of this work), }and we must discard it as a viable choice of $q$ for the purposes of this work. This is likely a consequence of the $q=2$ WST tending to over-emphasise spurious peaks in noisy observations.

\subsection{Detection significance}
\label{sec:snr}
We consider a kind of median expectation for the detection significance as follows. For each 2 GHz sideband $i$ of the COMAP Pathfinder or COMAP-ERA observation, we have a numerical estimate of the covariance matrix $\mathbfss{C}_{i;\alpha\beta}$ of the power spectrum or the WST coefficients (indexing $k$-bins or WST coefficients with $\alpha$ and $\beta$ here to avoid confusion with the indices $j$ and $j'$ used for the WST scales). We also have a median expectation value -- write it as $\avg{\Delta P(k)}$ or $\{\avg{\Delta s_1(j)},\avg{\Delta s_{20}(j)},\avg{\Delta s_2(j,j')}\}$, for example -- for the difference between the measured power spectrum or WST coefficient set and the expected statistic values from noise alone (i.e., in the absence of any CO signal). Then for each sideband $i$ we can effectively consider the Mahalanobis distance as a measure of the significance of the deviation of the measurement from noise, if we consider the expectation from noise alone to be the presumed mean of the distribution.

For the power spectrum,
\begin{equation}
    d_{M,i} = \sum_{\alpha,\beta}\avg{\Delta P(k_\alpha)}\mathbfss{C}_{i;\alpha\beta}^{-1}\avg{\Delta P(k_\beta)}.
\end{equation}
The total significance across all four sidebands is then
\begin{equation}
    \mathrm{\frac{S}{N}}\equiv\sqrt{\sum_i d_{M,i}}.
\end{equation}
If the power spectrum covariance matrix is diagonal for all $i$ such that $\mathbfss{C}_{i;\alpha\beta}=\sigma^2_{P(k);\alpha}\delta_{\alpha\beta}$, then the detection significance is simply $\sqrt{\sum_\alpha\avg{\Delta P(k_\alpha)}^2/\sigma_{P(k);\alpha}^2}$. Furthermore, since the total observed power spectrum is the sum of the signal and noise power spectra, $\avg{\Delta P(k_\alpha)}$ should be the same as the expected value of the signal power spectrum by itself. All of this is consistent with previous detection significance forecasts for COMAP.

The computation for the WST coefficients (raw or reduced) is analogous, except the sum is now over the range of coefficients rather than the $k$-bins of the power spectrum; we omit an explicit expression for the WST $\mathrm{S/N}$. As expected from the condition numbers for the covariance matrices, this calculation is outright impossible for the raw WST coefficients with $q=1$ due to the ill-conditioned covariance matrix, and although we show the results for the raw coefficients with $q=\{0.5,0.75,1.5\}$, major caveats apply as to their credibility. As for the shapeless WST coefficients, their covariance matrices are credibly invertible and therefore we show the results without the same caveats. We do not show results for the rescaled WST coefficients as they should be largely equivalent to those for the raw WST (modulo numerical instabilities).

As previously discussed, the covariance matrices could be estimated from either noise-only simulations or simulations that include both signal and noise. Given that the null hypothesis would be that our measurement is consistent with the distribution of $P(k)$ or WST measurements expected from noise alone, it arguably makes sense to calculate detection significance against the noise-only covariance. However, to quantify information content we should calculate detection significance against the signal-plus-noise covariance. We will present both calculations in~\autoref{tab:snr}.

\begin{table}
    \centering
    \begin{tabular}{r|c|c}
         \hline Statistic & \multicolumn{2}{c}{Computed $\mathrm{S/N}$ for:}\\
         & Pathfinder & COMAP-ERA \\\hline
         $P(k)$ & 11 (8) & 240 (21) \\
         Raw WST, $q=0.5$ & 30 (24) & 500 (96) \\
         Shapeless WST, $q=0.5$ & 12 (9) & 184 (34) \\
         Raw WST, $q=0.75$ & 30 (22) & 598 (87) \\
         Shapeless WST, $q=0.75$ & 12 (9) & 197 (32) \\
         Raw WST, $q=1.0$ & -- & -- \\
         Shapeless WST, $q=1.0$ & 12 (9) & 206 (30) \\
         Raw WST, $q=1.5$ & 39 (32) & 625 (61) \\
         Shapeless WST, $q=1.5$ & 13 (9) & 265 (30)% \\
         %Raw WST, $q=2.0$ & 1078 (915) & 5299 (1852) \\
         %Shapeless WST, $q=2.0$ & 12 (9) & 332 (33)
         \\\hline
    \end{tabular}
    \caption{Total expected detection significance of $P(k)$ and WST coefficients against noise covariance (values in parentheses are computed against signal-plus-noise covariance) for the COMAP Pathfinder and COMAP-ERA simulations. Dashes indicate that the detection significance could not be calculated due to a singular covariance matrix.}
    \label{tab:snr}
\end{table}

Regardless of whether we use the raw or shapeless WST coefficients, the WST detection significance is generallly in excess of the $P(k)$ detection significance. Based on the Pathfinder computations, the optimal value of $q$ would appear to be 1.5. This is still true when looking at COMAP-ERA detection significance against noise alone, which increases monotonically with $q$. However, for COMAP-ERA, the detection significance calculated against the total (signal-plus-noise) covariance matrices appears to decrease with greater $q$, perhaps a sign that information in CO underdensities becomes more apparent with lower noise levels. Overall, however, we will proceed with $q=1.5$, a configuration that still results in the WST apparently holding several times more information than the power spectrum.

\subsection{Parameter constraints}
\label{sec:fisherres}
Now fixing $q=1.5$ based on the above results, we consider Fisher forecasts for parameter constraints with the formalism established in~\autoref{sec:fisher}. The observables considered are the power spectrum $P(k)$ only, the shapeless WST coefficients, the rescaled WST coefficients, and the combination of $P(k)$ with the VID.

We now define the additional normalisations $N_1$ and $N_2$ required to fully define the rescaled coefficients $\tilde{S}_1$ and $\tilde{S}_2$. As might be gathered from~\autoref{fig:derivs_noisy}, the rescaling required to optimise covariance conditioning differs between the noise-dominated COMAP Pathfinder scenario and the significantly less noise-dominated COMAP-ERA scenario. For COMAP Pathfinder simulations we use the following definitions:
\begin{align}
    N_1(j,\ell) &= 1;\\
    N_2(j,j',\ell) &= \begin{cases}2^{J-j}&\text{if }\ell>0,\\1&\text{if }\ell=0.\end{cases}
\end{align}
For the COMAP-ERA scenario we instead use the following:
\begin{align}
    N_1(j,\ell) &= (\ell+1/2)2^{-j/2};\\
    N_2(j,j',\ell) &= (\ell+1/2).
\end{align}
Recall also that as part of defining the rescaled WST coefficients, we discard $\tilde{S}_2$ with $j'>j+1$ and $\ell=0$ due to very high correlation of these coefficients with $\tilde{S}_2(j,j'=j+1,\ell=0)$.

The above definitions of $N_1$ and $N_2$ lower the base-2 logarithm of the condition numbers of the covariance matrices from $\approx28$ to $\approx18$--19 for COMAP Pathfinder covariances, or a slightly higher range of 19--20 for COMAP-ERA. In all cases, our condition numbers are one order of magnitude better than the $10^7$ maximum imposed by~\cite{Eickenberg22}, which corresponds to $\approx2^{23}$ (just barely within 32-bit float precision).

\begin{table}
    \centering
    \begin{tabular}{l|c|c|c}
         \hline Statistic & \multicolumn{3}{c}{COMAP Pathfinder (COMAP-ERA) $1\sigma$ error for:} \\
         & $A$ & $C$ & $\sigma$ \\\hline
         $P(k)$ only & 0.130 (0.053) & 0.042 (0.025) & 0.119 (0.030)\\
         Shapeless WST & 0.126 (0.068) & 0.041 (0.016) & 0.095 (0.019)\\
         $P(k)$ \& VID & 0.100 (0.052) & 0.034 (0.015) & 0.091 (0.019)\\
         Rescaled WST & 0.014 (0.009) & 0.010 (0.008) & 0.015 (0.008)\\
         \emph{-- with $\ell>0$ only} & 0.021 (0.017) & 0.015 (0.010) & 0.027 (0.012)\\
         \hline
    \end{tabular}
    \caption{Marginalised parameter constraining power predicted by Fisher forecasts using various observables as described in the main text. The final row provides constraints from only $\tilde{S}_1(q,\ell>0)$ and $\tilde{S}_2(q,q',\ell>0)$.}
    \label{tab:fishersigmas}
\end{table}

\begin{figure}
    \centering
    \includegraphics[width=0.98\linewidth]{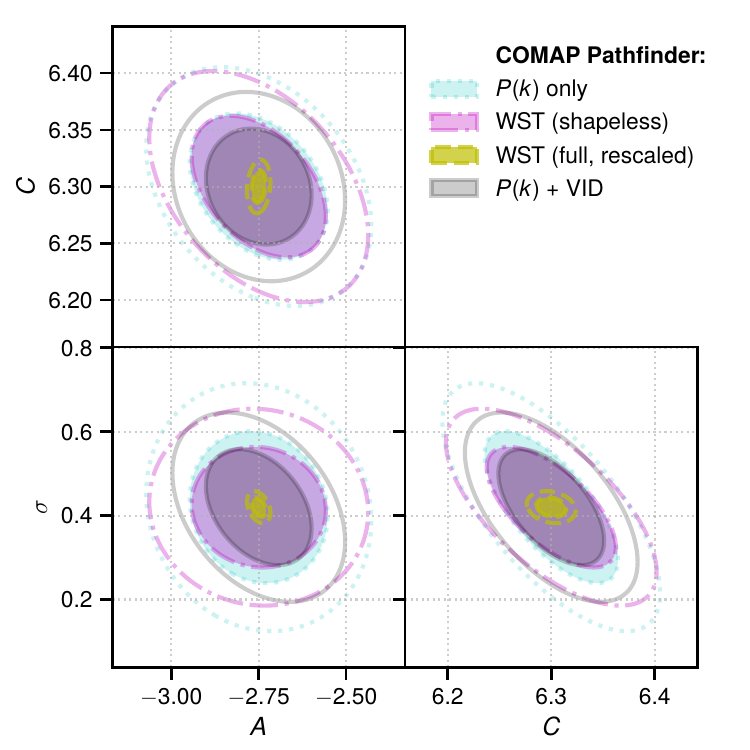}
    \includegraphics[width=0.98\linewidth]{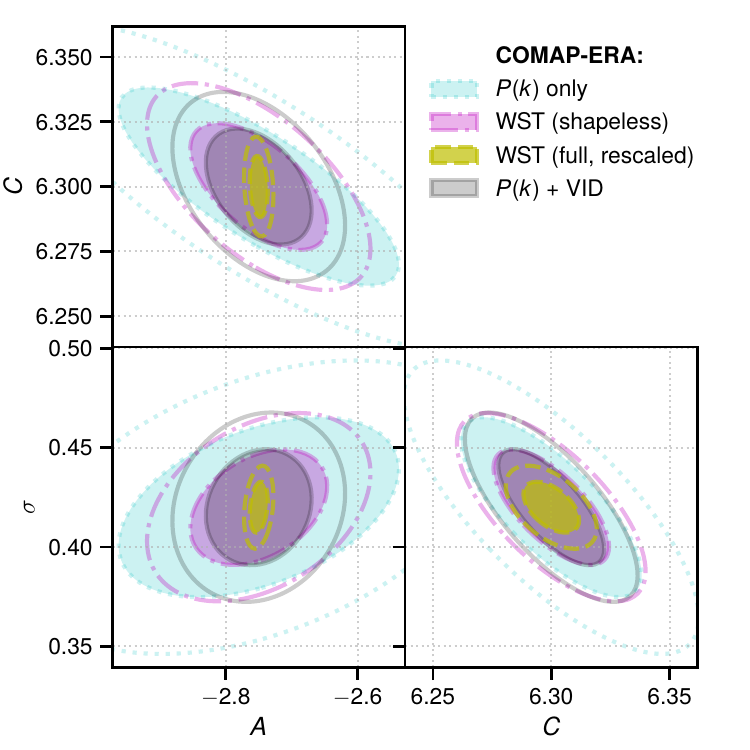}
    \caption{68\% and 95\% credibility ellipses from the Fisher forecasts for summary statistics (indicated in legends) from simulations of the COMAP Pathfinder (upper panels) and COMAP-ERA (lower panels).}
    \label{fig:fisher}
\end{figure}

We show the Fisher forecast results in two different ways: we show 68\% and 95\% credibility ellipses in~\autoref{fig:fisher}, and marginalised parameter errors in~\autoref{tab:fishersigmas}. These results in the COMAP Pathfinder case should not be directly compared to forecasts for the Pathfinder survey in~\cite{COMAPESV}. That work uses a Monte Carlo method and fully incorporates non-Gaussian and skewed parameter covariances in this way, which likely significant affects errors for $A$. We also fix $B$ and $M$ and thus it is natural to obtain tighter constraints on the remaining parameters, and this is likely a significant effect for $C$, which is highly degenerate with $M$ in the modelling and forecasts of~\cite{COMAPESV}.

Overall, we find that the WST has the fundamental sensitivity to allow parameter constraints finer by a factor of 2--6 compared to with the combination of $P(k)$ with the VID. However, note that the constraining power of the shapeless WST coefficients is comparable to the $P(k)$--VID combination. Also of note is the change in parameter constraints between the COMAP Pathfinder and COMAP-ERA scenarios. Whereas the improvement in marginalised parameter errors does not reach a factor of 2 for the rescaled WST coefficients, it reaches a factor of 2--4 for all other summary statistics used. This appears to match the relative changes in detection significance computed for $P(k)$ and the WST (both raw and shapeless coefficient sets) against the signal-plus-noise covariance in~\autoref{sec:snr}.

We also consider constraints with the full rescaled WST coefficients but only using the subset of 45 $\tilde{S}_1$ and $\tilde{S}_2$ values with $\ell>0$, i.e., only using the information content in the WST that derives from non-monopole local CO emission gradients. We find that the loss of $\ell=0$ information results in parameter constraints that are somewhat weaker than with the full rescaled WST coefficient set, but nonetheless stronger than those derived from the $P(k)$--VID combination.

\section{Discussion}
\label{sec:discussion}
\subsection{The shapeless WST coefficient set: key takeaways}

The reduction of the full WST coefficient to the shapeless set is somewhat sensible in a few different ways. The key motivation is the na\"{i}ve expectation that we do not gain signficant information from distinguishing between WST coefficients of different spherical harmonic multipoles $\ell$, and that the majority of information content in the WST is in variations of $s_1(j)$, $s_2(j,j')$, and $s_{20}(j)$ with scales $j$ and $j'$. We also observe an important numerical advantage -- likely helped by reduced dimensionality -- that correlation coefficient and covariance matrices are much better conditioned than for the raw WST, and without significant fine-tuning as was required for the rescaled WST coefficients.

However, subsequent calculations of detection significance and parameter constraints in the latter half of~\autoref{sec:results} clearly show that the shapeless WST coefficients do not capture the full information in the WST. In fact, the similarity of constraining power between the $P(k)$--VID combination and the shapeless WST coefficients should be striking. Neither $P(k)$ nor the VID will be sensitive to the `shape' of CO fluctuations -- the addition of the VID provides non-Gaussian but shape-free information. Therefore, it is natural that we do no better than the $P(k)$--VID combination once we discard variations with $\ell$ and thus morphological information from the WST.

\subsection{The full WST coefficient set: practical considerations}

The fundamental sensitivity of the WST in a COMAP Pathfinder or COMAP-ERA observational scenario is extremely promising. In principle, deviations of the WST from the noise-only baseline can be detected at considerably higher significance than the CO $P(k)$, and parameter constraints with the full rescaled WST coefficients are correspondingly tighter by an order of magnitude in the COMAP Pathfinder case. They are only tighter by a factor of 3--5 in COMAP-ERA simulations, but this is likely conservative as making use of additional values of the density-weighting exponent $q$ could further improve WST-based constraints. With lower noise in the COMAP-ERA scenario, CO voids should become more readily apparent, and evaluations of WST coefficients with $q=0.5$ or $q=0.75$ could provide additional information content. Indeed,~\cite{Eickenberg22} are apparently able to use two or even five different values of $q$ at once, although their work only calculates summary statistics including the WST on noiseless matter density fields.

However, before even considering using multiple values of $q$ at once, LIM experiments should consider whether they can make practical use of the WST with even just one value of $q$. Our results are strongly contingent on the ability to understand the interaction between signal and noise in the WST. Our use of a beam-smoothed, high-pass filtered CO field summed with a uniform Gaussian field mirrors the approximate input for a pseudo-$P(k)$ analysis as undertaken by COMAP~\citep{Ihle21}. However, non-uniform noise residuals from removal of systematics and foregrounds may eventually manifest above uniform noise with greater integration times. The interaction between the CO signal and uniform noise is already quite complex, and only further complexity can arise from systematics like standing waves in the instrument or other pathological variations in the noise level across frequency, making the WST analysis more tenuous in practical terms even beyond mode mixing considerations.

Making matters even more tenuous is the fact that the WST does not have an established extension to cross-correlations as the power spectrum does. The pseudo-cross $P(k)$ analysis of~\cite{Ihle21} is made more robust by the fact that the result is not generated from a single co-added map, but from cross-correlation power spectra between various pairs of pseudo-temperature maps generated from different data splits. This methodology intrinsically rejects disjoint systematics and uncorrelated noise between the splits, thus improving the estimate of the underlying CO power spectrum. Such cross-correlation power spectra are mathematically well motivated since the auto-correlation is merely a special case of the cross-correlation spectrum. Meanwhile, the key insight behind generalisation of the VID to cross-distribution statistics~\citep{Breysse19} is the straightforward convolution of signal and noise distributions and the ability to deconvolve uncorrelated noise out of distribution shapes in Fourier space. Due to the nonlinear nature of the WST, gaining similar insights and defining cross-field extensions may prove difficult.

The ease with which $P(k)$ by contrast can be applied to cross-correlations is key not only for mitigation of systematics within a single experiment, but also for cross-correlations between independent measurements of the same cosmic web for multi-tracer astrophysics on cosmological scales. The COMAP survey has been designed for cross-correlations with spectroscopic galaxy surveys, in particular the Hobby-Eberly Telescope Dark Energy eXperiment~\citep[HETDEX;][]{HETDEX2008,HETDEX2021inst,HETDEX2021} surveying Lyman-alpha emitters (LAE) at redshifts overlapping with COMAP target redshifts. Fisher forecasts already predict approximately a factor-of-two improvement in determination of the mean CO temperature--bias product with the CO--LAE cross power spectrum versus the CO auto $P(k)$. Certainly the CO `auto' WST promises even further improvement but at the cost of reduced robustness compared to a pseudo-cross $P(k)$ analysis and no readily apparent path to a cross-field extension.

On top of all of this, the WST covariance does not appear to be analytically understood, given that it is numerically estimated in every work that has been cited in this paper (and in this paper as well). The nonlinear, multi-scale nature of the WST appears to be both an advantage in the resulting rich morphological and non-Gaussian information content, and a severe disadvantage in complicating covariance considerations and thus affecting the ability to extract desired constraints from all of that information content.

\section{Conclusions}
\label{sec:conclusions}

With our results, we have answered the questions that we set out in~\hyperref[sec:intro]{the Introduction} (reproduced below):
\begin{itemize}
    \item \emph{What is the detectability of cosmological CO emission in the WST coefficients derived from a simulated COMAP observation?} The fundamental sensitivity of the COMAP Pathfinder should allow for the detection of deviations of the full WST coefficient set from noise. The predicted overall signal-to-noise ratio of 30--40 with the Pathfinder survey is several times what is expected for the CO $P(k)$. The detectability increases even further with the lower noise level expected from COMAP-ERA, although information content only increases by the same relative amount as for $P(k)$.
    \item \emph{Do parameter constraints from these coefficients improve significantly over the previously studied combination of $P(k)$ with the VID?} Fisher forecasts suggest that the full (rescaled) WST coefficient set allows for parameter constraints at least 2--5 times finer than the $P(k)$--VID combination. This potential advantage is specifically due to the additional shape information provided across the basis of solid harmonic wavelets with varying $\ell$, and could potentially be magnified by use of different values of the density-weighting exponent $q$.
\end{itemize}

These exploratory results are highly encouraging and seem to corroborate that wavelets are a more optimal basis than Fourier modes for probing LSS, even as traced by CO. However, as we have discussed above, the covariance of the WST still needs to be better understood.\replacedtwo{ This}{ Illustrating this need is our discovery that for the solid harmonic WST applied to simulated noisy data, the covariance matrix of the full coefficient set becomes singular as the WST hyperparameter $q$ goes to 1. For the forecasts in this work, appropriate choices of $q$ and rescaled or reduced coefficient sets kept covariances well-conditioned and avoided such pathological behaviour. Nonetheless, this singularity -- which appears to be newly found in this work -- has clear, important implications even outside the context of CO LIM. A full understanding of this and other aspects of WST covariances} is vital to any real-world likelihood analyses that would make use of the WST, given strong inherent correlations between WST coefficients and given the complex interaction of signal and noise.

Perhaps even more important is the design of a cross-correlation extension of the WST, without which it is not possible to design an analogue to a pseudo-cross $P(k)$ analysis that rejects uncorrelated systematics and noise in LIM data. As previously mentioned, \cite{Breysse19} already formulate a VID extension to cross-correlate galaxy counts with 21 cm intensity measurements, and other work is already extending this to make use of joint PDFs between different line-intensity fields (Breysse et al., in prep.). Hopefully, similar future extensions to the WST will allow robust auto- and cross-correlation measurements of the morphological and non-Gaussian information content that seems uniquely accessible to a WST-type LIM analysis.

\section*{Acknowledgements}

Thanks to Dick Bond, Jonathan Braden, and Patrick Breysse for various wavelet- and/or WST-related conversations leading up to and during the writing of this paper. Thanks also to Veronica Stewart for a conversation that helped clarify the physical chemistry context underlying the original development of the solid harmonic WST by~\cite{Eickenberg18}.

It is a pleasure to acknowledge the COMAP collaboration for allowing use of approximate noise levels and transfer functions to inform simulation parameters. Many thanks to George Stein for running and making available the original peak-patch simulations for~\cite{Ihle19} that are reused in this work (and in~\citealt{COMAPESV}).\added{ Thanks also go to an anonymous referee whose constructive comments improved this manuscript.}

Research in Canada is supported by NSERC and CIFAR. Parts of these computations were performed on the GPC supercomputer at the SciNet HPC Consortium, and on workstations hosted at CITA. SciNet is funded by: the Canada Foundation for Innovation under the auspices of Compute Canada; the Government of Ontario; Ontario Research Fund -- Research Excellence; and the University of Toronto.

DTC is supported by a CITA/Dunlap Institute postdoctoral fellowship. The Dunlap Institute is funded through an endowment established by the David Dunlap family and the University of Toronto. The University of Toronto operates on the traditional land of the Huron-Wendat, the Seneca, and most recently, the Mississaugas of the Credit River; DTC is grateful to have the opportunity to work on this land. This research made use of Astropy,\footnote{http://www.astropy.org} a community-developed core Python package for astronomy \citep{astropy:2013, astropy:2018}. This research also made use of NASA's Astrophysics Data System Bibliographic Services. 

%%%%%%%%%%%%%%%%%%%%%%%%%%%%%%%%%%%%%%%%%%%%%%%%%%
\section*{Data Availability}

The data underlying this article will be shared on reasonable request to the author.



\bibliographystyle{mnras}
\bibliography{ms}

%%%%%%%%%%%%%%%%%%%%%%%%%%%%%%%%%%%%%%%%%%%%%%%%%%

%%%%%%%%%%%%%%%%% APPENDICES %%%%%%%%%%%%%%%%%%%%%

%\appendix

%%%%%%%%%%%%%%%%%%%%%%%%%%%%%%%%%%%%%%%%%%%%%%%%%%

% Don't change these lines
\bsp	% typesetting comment
\label{lastpage}
\end{document}